\documentclass[11pt,twoside,dvips,a4paper]{article}
\usepackage{graphicx,color}
\usepackage{times}
\usepackage{booktabs}
\usepackage{cite}
\usepackage{epsfig}
\usepackage{color}
\usepackage{a4wide}

\newcommand{\mrm}[1]{\mathrm{#1}}

\newcommand{\ttt}[1]{\texttt{#1}}
\newcommand{\pT}[1]{\ensuremath{p_{\perp#1}}}
\newcommand{\TeV}{\,\mbox{Te\kern-0.2exV}}
\newcommand{\GeV}{\,\mbox{Ge\kern-0.2exV}}
\newcommand{\MeV}{\,\mbox{Me\kern-0.2exV}}
\newcommand{\keV}{\,\mbox{ke\kern-0.2exV}}
\newcommand{\eV}{\,\mbox{e\kern-0.2exV}}

\begin{document}
\vspace*{-2.0cm}\noindent\begin{minipage}{\textwidth}
\flushright
FERMILAB-CONF-09-113-T
\end{minipage}\vspace*{0.75cm}
\begin{center}
\Large{\bf The ``Perugia'' Tunes}\\[8mm]
{\normalsize {\bf P.~Skands} (skands@fnal.gov)\\
{\sl Theoretical Physics, Fermilab, MS106, Box 500, Batavia
  IL-60510, USA}}\\[8mm]
\end{center}

\begin{abstract}
We present 7 new tunes of the $p_\perp$-ordered shower and
underlying-event model in \textsc{Pythia} 6.4. These ``Perugia'' tunes 
update and supersede the older ``S0'' family. The new tunes 
include the updated LEP fragmentation and flavour parameters reported
on by H.~Hoeth at this workshop \cite{HoethProc}. The hadron-collider
specific parameters were then retuned (manually) using Tevatron
min-bias data from 630, 
1800, and 1960 GeV, Tevatron Drell-Yan data at 1800 and 1960 GeV, 
as well as SPS min-bias data at 200, 540, and 900 GeV. In addition to
the central parameter set, related tunes exploring systematically
soft, hard, parton density, and color structure variations are
included. Based on these variations, a best-guess prediction of 
the charged track multiplicity in inelastic, nondiffractive
minimum-bias events at the LHC is made. 
\end{abstract}

\section{Introduction}

Perturbative calculations of collider observables rely
on two important prerequisites: factorisation and
infrared safety. These are the tools that permit us to relate
the calculations to detector-level measured quantities, up to
corrections of 
known dimensionality, which can then be suppressed (or enhanced) by 
appropriate choices of the dimensionful scales appearing in
the poblem. However, this approach does limit us to 
consider only a predefined class of observables, at a limited
precision set by the aforementioned scales. In the context of
the underlying event, say, 
we are faced with the fact that we do not (yet) have
factorisation theorems for this component, while at the same time
acknowledging that not all collider measurements can be made
insensitive to it at a level comparable to the achievable experimental
precision. And when considering observables such as track
multiplicities, hadronisation corrections, or even short-distance
resonance masses if the precision required is very high, we are confronted
with quantities which may be experimentally well measured but which
are  explicitly sensitive to infrared physics. 

Let us
begin with factorisation. When applicable, factorisation allows us to
subdivide the calculation of an observable (regardless of whether it
is infrared safe or not) into a perturbatively calculable
short-distance part and a universal long-distance part, the latter of
which may be modeled and constrained by fits to data. However, in the
context of hadron collisions the conceptual 
separation into ``hard-scattering'' and ``underlying-event''
components is not necessarily equivalent to a clean separation in
terms of ``hardness'' (or perhaps more properly formation time), 
since what is labeled the ``underlying
event'' may contain short-distance physics of its own. Indeed, from
ISR energies \cite{Akesson:1986iv} through the SPS
\cite{ua1minijets,Alitti:1991rd} to  
the Tevatron
\cite{Abe:1993rv,Abe:1997bp,Abe:1997xk,Abazov:2002mr,dmitry}, and even
in photoproduction at HERA
\cite{Gwenlan:2002st}, we see evidence of (perturbative)
``minijets'' in the underlying event, beyond what bremsstrahlung alone
appears to be able to account for. It would therefore
seem apparent that a universal modeling of the underlying event 
must include at least some degree of correlation between the
hard-scattering and underlying-event components. It is in this spirit
that the concept of ``interleaved evolution'' \cite{Sjostrand:2004ef} 
was developed as the cornerstone of the $\pT{}$-ordered models
\cite{Sjostrand:2004pf,Sjostrand:2004ef} in both 
\textsc{Pythia}~6~\cite {Sjostrand:2006za} and, more recently,
\textsc{Pythia}~8~\cite{Sjostrand:2007gs}.  

The second tool, infrared safety, provides us
with a class of observables which are insensitive to the details of
the long-distance physics. This works up to corrections of order the
long-distance scale divided by the short-distance scale,
$Q_{\mathrm{IR}}^2/Q_{\mathrm{UV}}^2$, where $Q_\mathrm{UV}$ denotes
a generic hard scale in the problem and $Q_\mrm{IR} \sim
\Lambda_\mrm{QCD} \sim \mathcal{O}(\mrm{1\ GeV})$.  Since
$Q_{\mathrm{IR}}/Q_{\mathrm{UV}}\to 0$ for large $Q_{\mathrm{UV}}$, such
observables ``decouple'' from the infrared physics as long as all
relevant scales are $\gg Q_{\mathrm{IR}}$. Only if we require a
precision that begins to approach $Q_{\mathrm{IR}}$ should we begin to
worry about non-perturbative effects for such observables. Infrared
sensitive  
quantities, on the other hand, 
contain logarithms $\log^n(Q_{\mathrm{UV}}^2/Q_{\mathrm{IR}}^2)$ which grow
increasingly large as $Q_{\mathrm{IR}}/Q_{\mathrm{UV}}\to 0$.  
As an example, consider particle or track 
multiplicities; in the absence of nontrivial infrared
effects, the number of partons that would be
mapped to hadrons in a na\"ive local-parton-hadron-duality
\cite{Azimov:1984np} picture depends logarithmically on the infrared
cutoff. 

Min-bias/UE physics can therefore be perceived of as offering 
an ideal lab for studying nonfactorized and nonperturbative
phenomena with the highest possible statistics, giving crucial tests
of our ability to model and understand these ubiquitous components. 
As a beneficial side effect, the improved models and tunes that result
from this effort are important ingredients in the modeling of 
high-$\pT{}$ physics, in the context of which the underlying event and
nonperturbative effects  furnish a nontrivial ``haze'' into  which
the high-$\pT{}$ physics is embedded.  

As part of the effort to spur more interplay between theorists and
experimentalists in this field, we here report on a new set of tunes 
of the $\pT{}$-ordered \textsc{Pythia} framework, 
which update and supersede the older ``S0''
family of tunes. The new tunes have been made available via the routine
PYTUNE starting from \textsc{Pythia} version 6.4.20.

We have here focused in particular on the energy
scaling from lower energies towards the LHC and on attempting to provide
at least some form of systematic uncertainty estimates, in the form of
a small number of alternate parameter sets that represent systematic
variations in some of the main tune parameters

We also present a few distributions that carry
interesting and complementary information about the underlying
physics, updating and complementing those contained in
\cite{Skands:2007zz}.  For brevity, this text only includes a
representative selection, with more results available on the web
\cite{lhplots}. 

The main point is that, while each plot represents a
complicated cocktail of physics effects, such that any sufficiently
general model presumably could be tuned to give an acceptable
description observable by observable, it is very difficult 
to simultaneously describe the entire set. The real game 
is therefore not to study one distribution in detail, but to study the
degree of simultaneous agreement or disagreement over many, mutually
complementary, distributions. 

We have tuned the Monte Carlo in four consecutive steps:
\begin{enumerate}
\item Final-State Radiation (FSR) and Hadronisation (HAD): 
 using LEP data, tuned by Professor \cite{HoethProc,Buckley:2009ad}.  
\item Initial-State Radiation (ISR) and Primordial $k_T$: using 
  the Drell-Yan $\pT{}$ spectrum at 1800 and
  1960 GeV, as measured by CDF \cite{Affolder:1999jh} and D\O\ 
  \cite{:2007nt}, respectively. We treat the data as fully corrected
  for photon bremsstrahlung effects in this case, i.e., we compare the
  measured points to the Monte Carlo distribution of the original $Z$
  boson.  We believe this to be
 reasonably close to the definition used for the data points in both
 the CDF and D\O\ studies.
\item Underlying Event (UE) and Beam Remnants (BR): using 
  $N_\mrm{ch}$ \cite{Acosta:2001rm}, 
$dN_\mrm{ch}/d\pT{}$ \cite{Abe:1988yu}, and
  $\left<\pT{}\right>(N_\mrm{ch})$ \cite{moggi} in min-bias events  
at 1800 and 1960 GeV, as measured by CDF. Note that the $N_\mrm{ch}$ spectrum 
extending down to zero $\pT{}$ measured by the E735 Collaboration at
1800 GeV \cite{Alexopoulos:1998bi} was left out of the tuning, 
since we were not able to consolidate this measurement with the rest
of the data. We do not know
whether this is due to intrinsic limitations in the modeling or to a
misinterpretation on our part of the measured result. 
\item Energy Scaling: using $N_\mrm{ch}$ in min-bias events 
at 200, 540, and 900 GeV, as measured by
UA5~\cite{Alner:1987wb,Ansorge:1988kn}, and at 630 and 1800 GeV, 
as measured by CDF~\cite{Acosta:2001rm}. Note that 
we include neither elastic nor diffractive Monte Carlo events in any
of our comparisons, which could 
affect the validity of the modeling for the first few
bins in multiplicity. We therefore assigned less
importance to these bins when doing the tunes. The last two steps were 
iterated a few times.
\end{enumerate}
Note that the clean separation between the first and second points
assumes jet universality, i.e., that a $Z^0$, for instance,
fragments in the same way at a hadron collider as it did at LEP. This
is not an unreasonable first assumption, but it is still important to check it
explicitly, e.g., by measuring strange to unstrange
particle production ratios, vector to pseudoscalar meson ratios,
and/or baryon to meson ratios \emph{in situ} at hadron colliders.

Note also that we do not include any explicit ``underlying-event''
observables here. Instead, we rely on the large-multiplicity tail of
minimum-bias events to mimic the underlying event. 
A similar procedure was followed for the older ``S0'' tune
\cite{Sandhoff:2005jh,Skands:2007zg}, which 
turned out to give a very good simultaneous description of both
minimum-bias and underlying-event physics at the Tevatron, 
despite only having been tuned on minimum-bias data
there\footnote{Note: when
extrapolating to other energies, the alternative scaling represented
by ``S0A'' appears to be preferred over the default scaling used in
``S0''.}.  Conversely, Rick Field's ``Tune A'' 
\cite{tunea,Field:2005sa,Field:2005yw,Albrow:2006rt} was
originally only tuned on underlying-event data, but turned out to give
a very good simultaneous description of minimum-bias physics. We
perceive of this as good, if circumstantial, evidence of the
universal properties of the \textsc{Pythia} modeling. 

Additional important quantities to consider for further validation
(and eventually tuning, e.g., in the Professor framework), 
would be observables 
involving explicit jet reconstruction and explicit underlying-event 
observables in leading-jet, dijet, jet + photon, and Drell-Yan
events. Some of these have already been included in the Professor
framework, see \cite{HoethProc,Buckley:2009ad}. See also the 
underlying-event sections in the HERA-and-the-LHC
\cite{Alekhin:2005dx}, Tevatron-for-LHC \cite{Albrow:2006rt}, and Les
Houches write-ups \cite{Buttar:2008jx}.

\section{Main Features of the Perugia Tunes}

In comparison with tunes of the old  (\textsc{Pythia} 6.2) 
framework \cite{Sjostrand:1987su}, such as Tune 
A~\cite{tunea,Field:2005sa,Field:2005yw,Albrow:2006rt}, all
tunes of the new framework share a few common features. Let us first
describe those, with plots to illustrate each point, 
and then turn to the properties of the individual tunes. 

\begin{figure}[t]
\begin{center}\hspace*{-2mm}
\includegraphics*[scale=0.34]{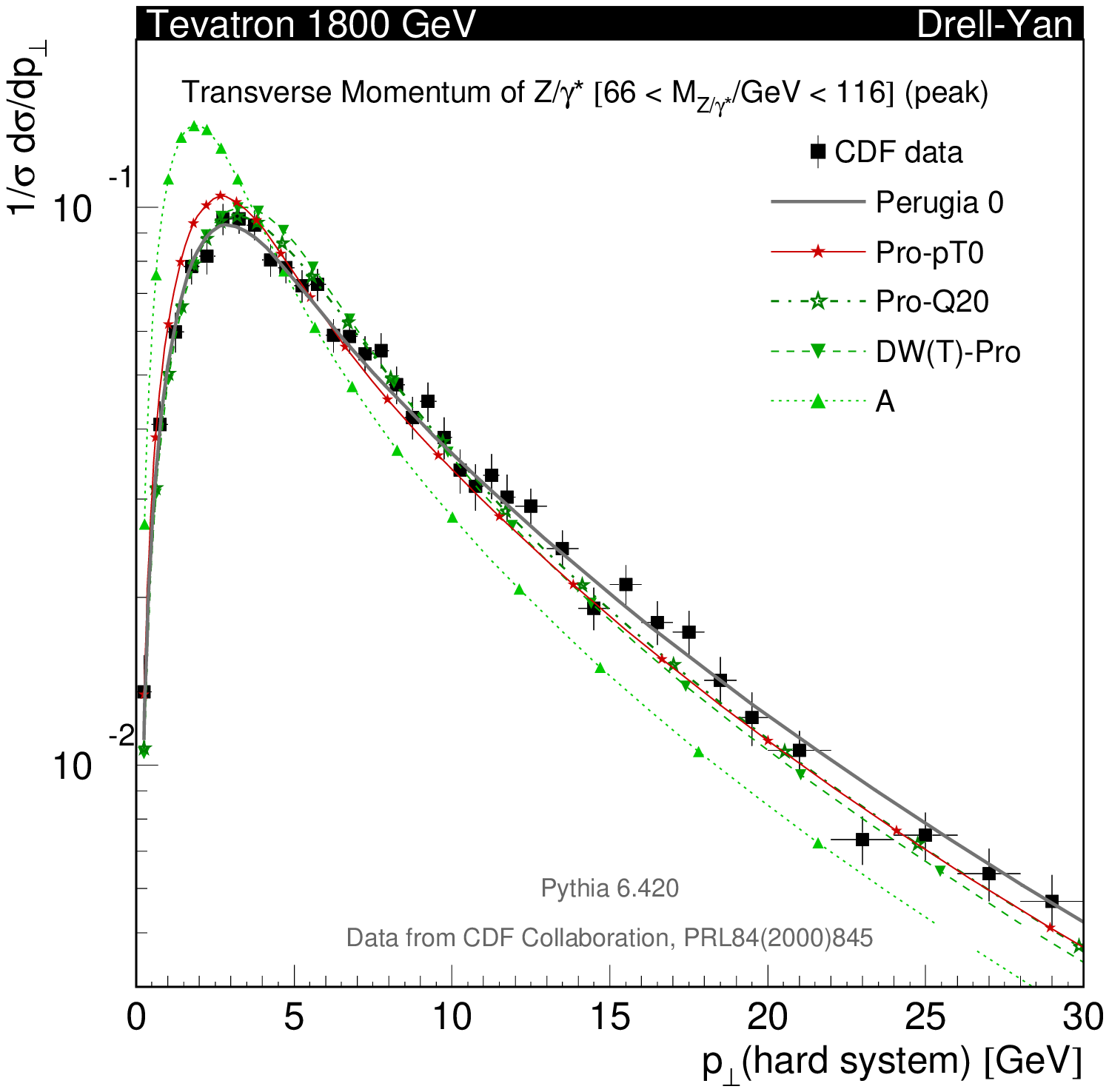}\hspace*{-5mm}
\includegraphics*[scale=0.34]{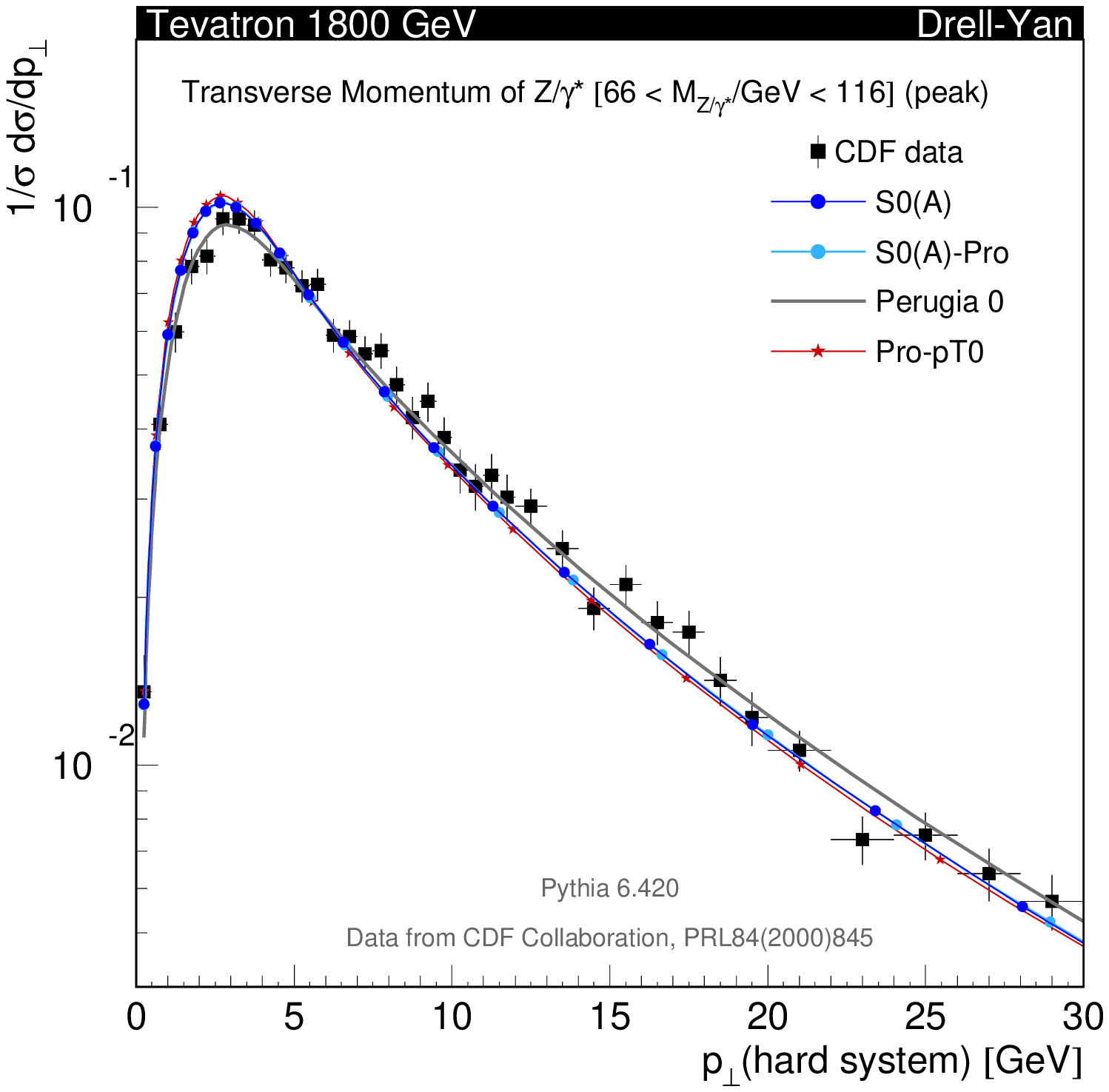}\hspace*{-5mm}
\includegraphics*[scale=0.34]{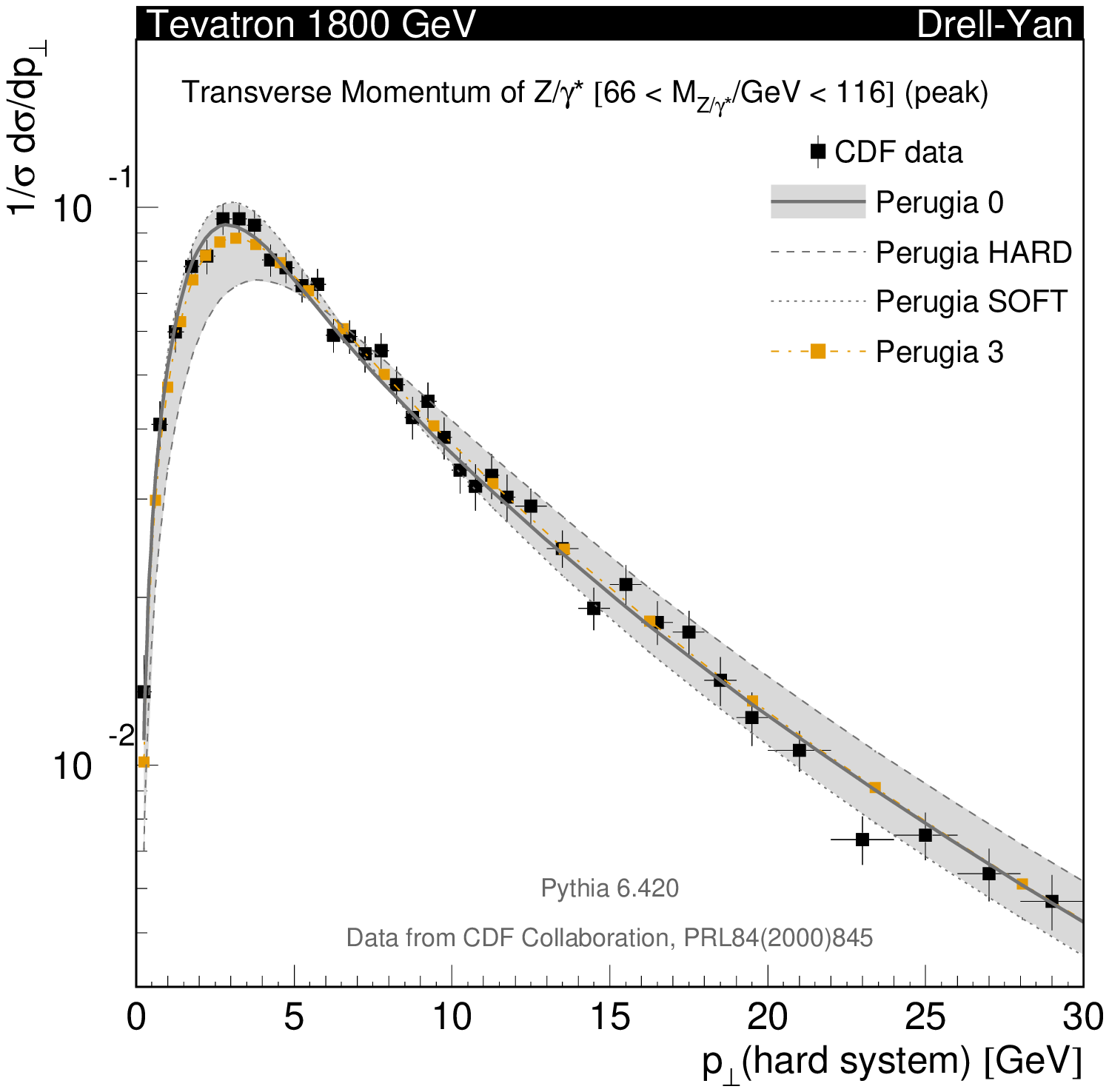}\hspace*{-8mm}\\[-4mm]
\caption{Comparisons to the CDF Run I measurement of
 the $\pT{}$ of Drell-Yan pairs \cite{Affolder:1999jh}. 
 {\sl Left:} a representative
 selection of models.  {\sl Center:} different tunes of the new
 framework. {\sl Right:} the range spanned by the 
main Perugia variations. 
Comparisons to the D\O\ Run II measurement \cite{:2007nt} and 
results with more tunes can be found at \cite{lhplots}.
Note that the Monte
 Carlo curves shown are for the $\pT{}$ of the original boson rather
 than of the lepton pair after (QED) showering.
\label{fig:tevatronDY}}
\end{center}
\end{figure}
First of all, the new
$\pT{}$-ordered showers \cite{Sjostrand:2004ef} 
employ a dipole-style recoil model, which
appears to make it very easy to obtain a good agreement with, e.g.,
the Drell-Yan $\pT{}$ spectrum. In the old model with default
settings, the Drell-Yan spectrum is only well described if FSR off ISR
jets is switched off. When switching this back on, which is of course 
necessary to
obtain the desired perturbative broadening of the ISR jets, 
the old shower kinematics work in such a way that each FSR emission
off a final-state parton from ISR 
effectively removes $\pT{}$ from the $Z$ boson,
shifting the spectrum towards lower values. This causes any tune of
the old \textsc{Pythia} framework with default ISR settings --- such as
Tune A or the ATLAS DC2/``Rome'' tune --- 
to predict a too narrow spectrum for the Drell-Yan $\pT{}$
distribution, as illustrated in fig.~\ref{fig:tevatronDY}. 

To re-establish agreement with the measured spectrum without changing
the recoil kinematics, the total amount of ISR in the old model 
had to be increased. This was done by choosing extremely low values of the
renormalisation scale (and hence large $\alpha_s$ values) for ISR
(tunes DW-Pro and Pro-Q20 in fig.~\ref{fig:tevatronDY}). 
While this nominally works, the whole business does smell faintly of
fixing one problem by introducing another and hence 
the default in \textsc{Pythia} has remained the unmodified Tune A,
at the price of retaining the poor agreement with the Drell-Yan
spectrum.  

In the new $\pT{}$-ordered
showers~\cite{Sjostrand:2004ef}, however, FSR off ISR is treated
within individual QCD dipoles and does not affect the Drell-Yan
$\pT{}$. This appears to make the spectrum come out generically much
closer to the data. The only change from the standard
$\alpha_s(\pT{})$ choice used in the S0 family of tunes was thus 
switching to the so-called CMW choice \cite{Catani:1990rr} 
for $\Lambda_{\mrm{QCD}}$ for ISR in the Perugia tunes, rather than the 
$\overline{\mrm{MS}}$ value used previously, similarly to what is done in
\textsc{Herwig} \cite{Corcella:2000bw,Bahr:2008pv}. 
 The effect of this relatively small 
change can be seen by comparing S0(A), which uses the
$\overline{\mrm{MS}}$ value, to Perugia 0 in the middle plot
on fig.~\ref{fig:tevatronDY}. The extremal curves on the right plot
are obtained by using $\alpha^{\mrm{CMW}}_s(\frac12\pT{})$ (HARD) 
and $\alpha^{\overline{\mrm{MS}}}_s(\sqrt{2}\pT{})$ (SOFT).

\begin{figure}[t]
\begin{center}
\hspace*{-2mm}
\includegraphics*[scale=0.34]{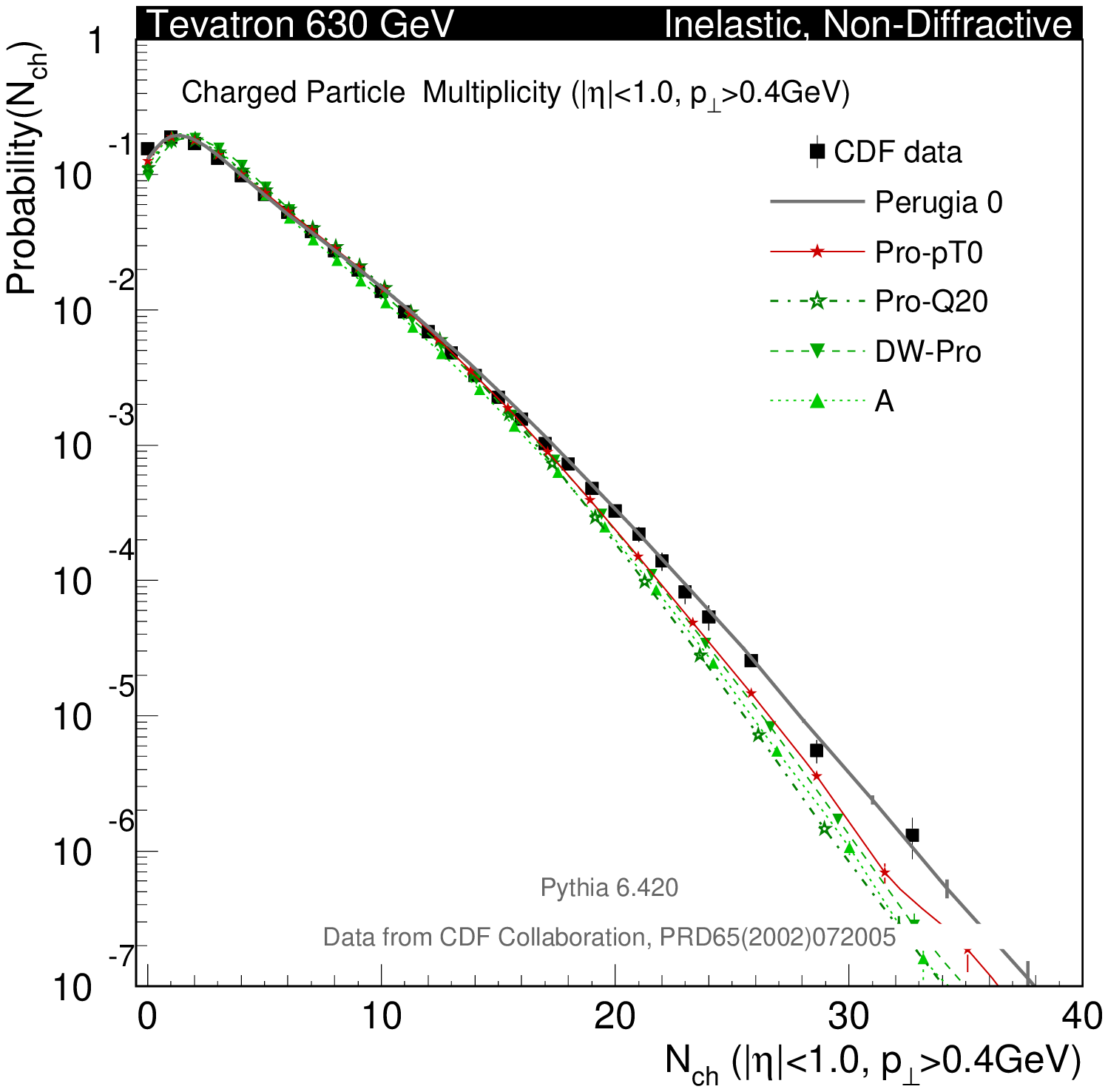}\hspace*{-5mm}
\includegraphics*[scale=0.34]{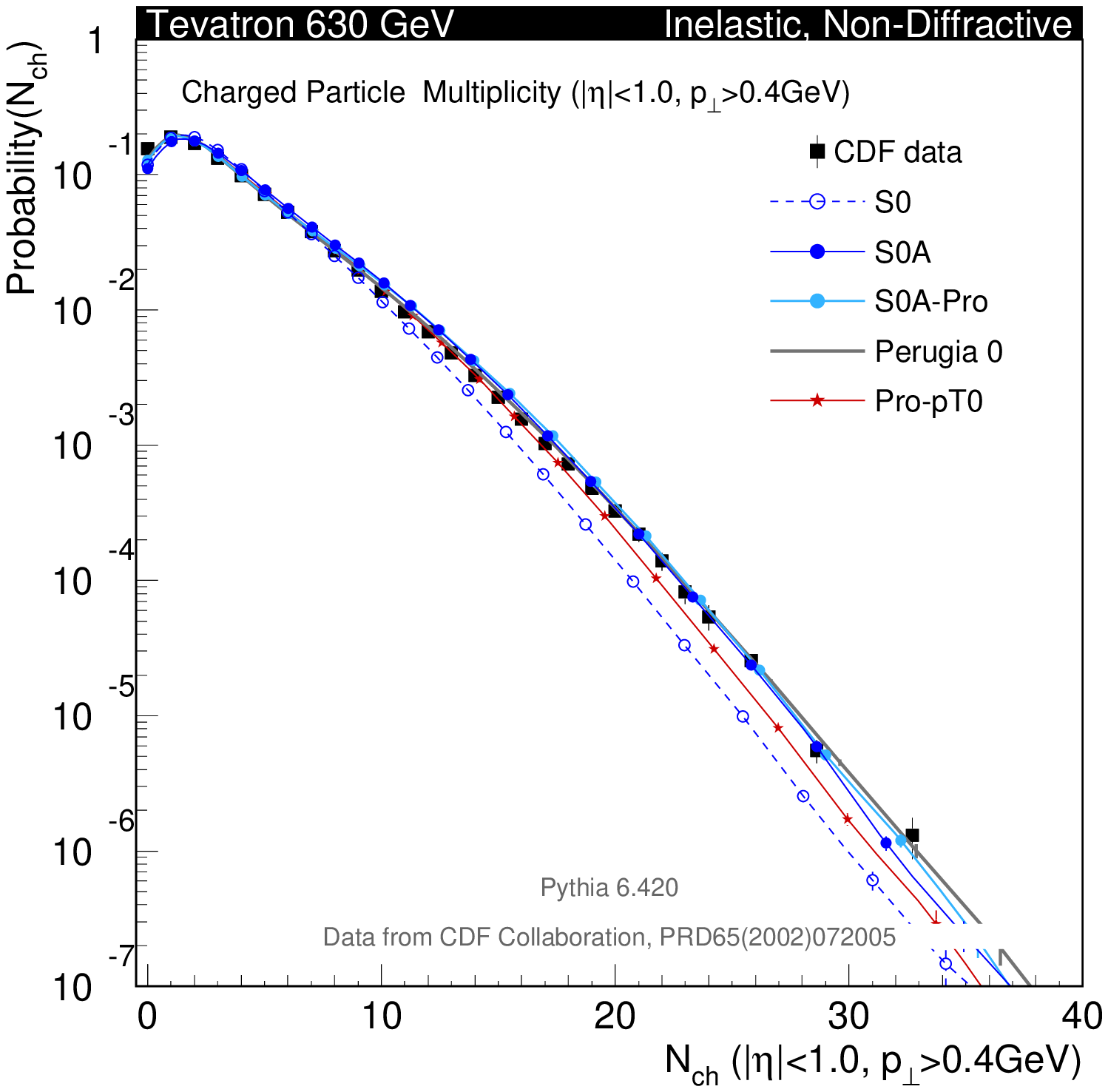}\hspace*{-5mm}
\includegraphics*[scale=0.34]{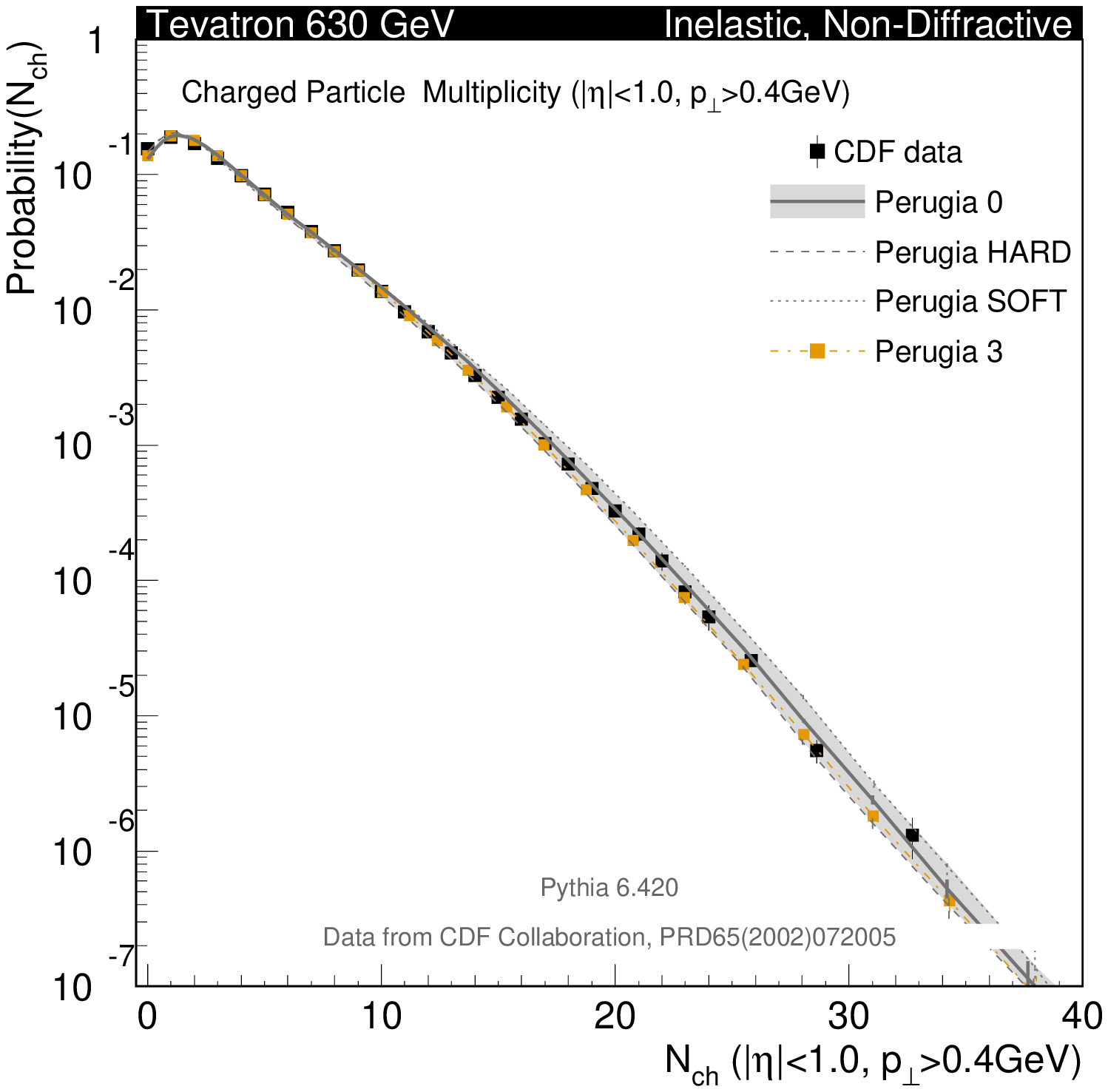}\hspace*{-8mm}\\[-5mm]
\hspace*{-2mm}
\includegraphics*[scale=0.34]{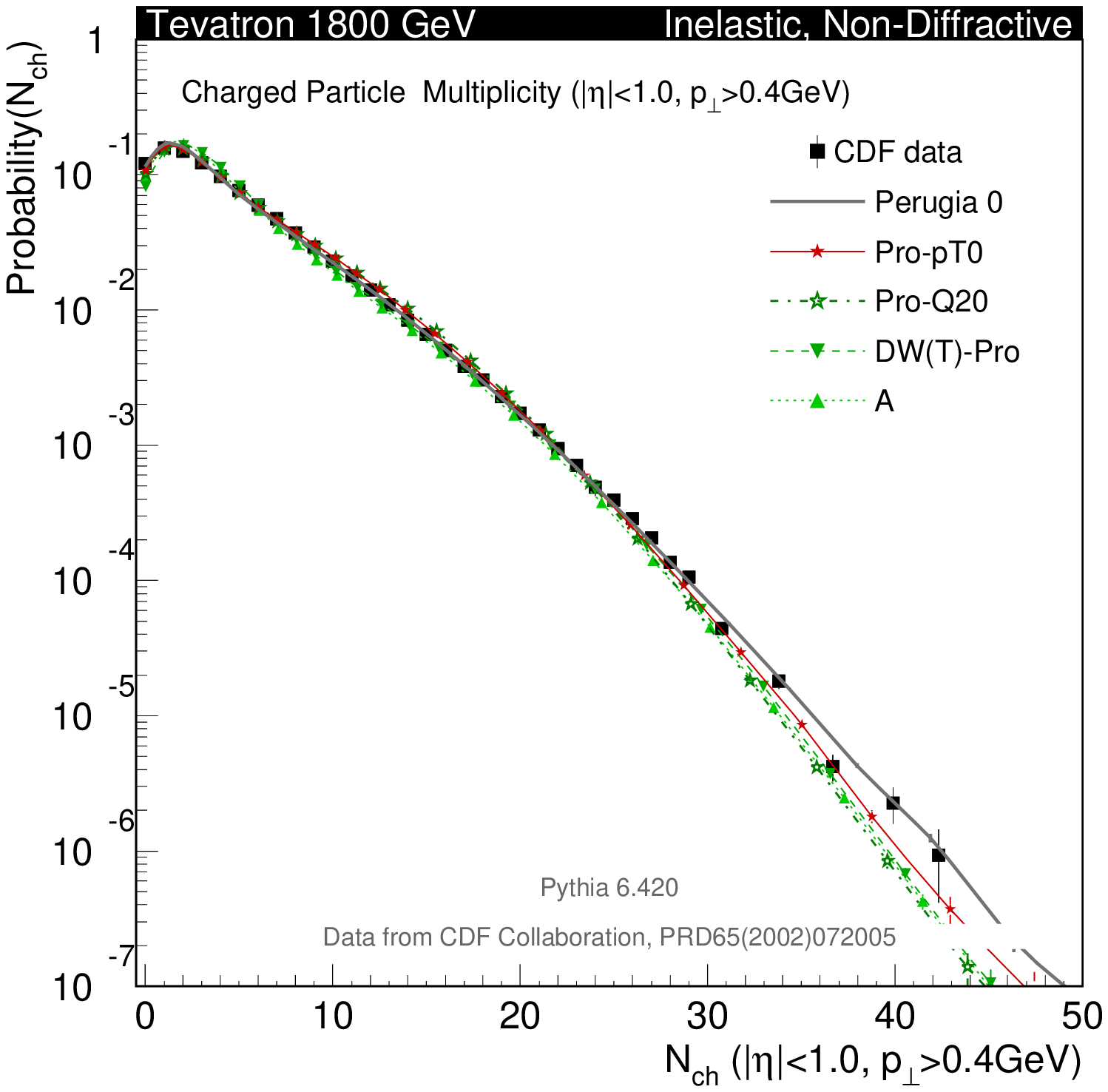}\hspace*{-5mm}
\includegraphics*[scale=0.34]{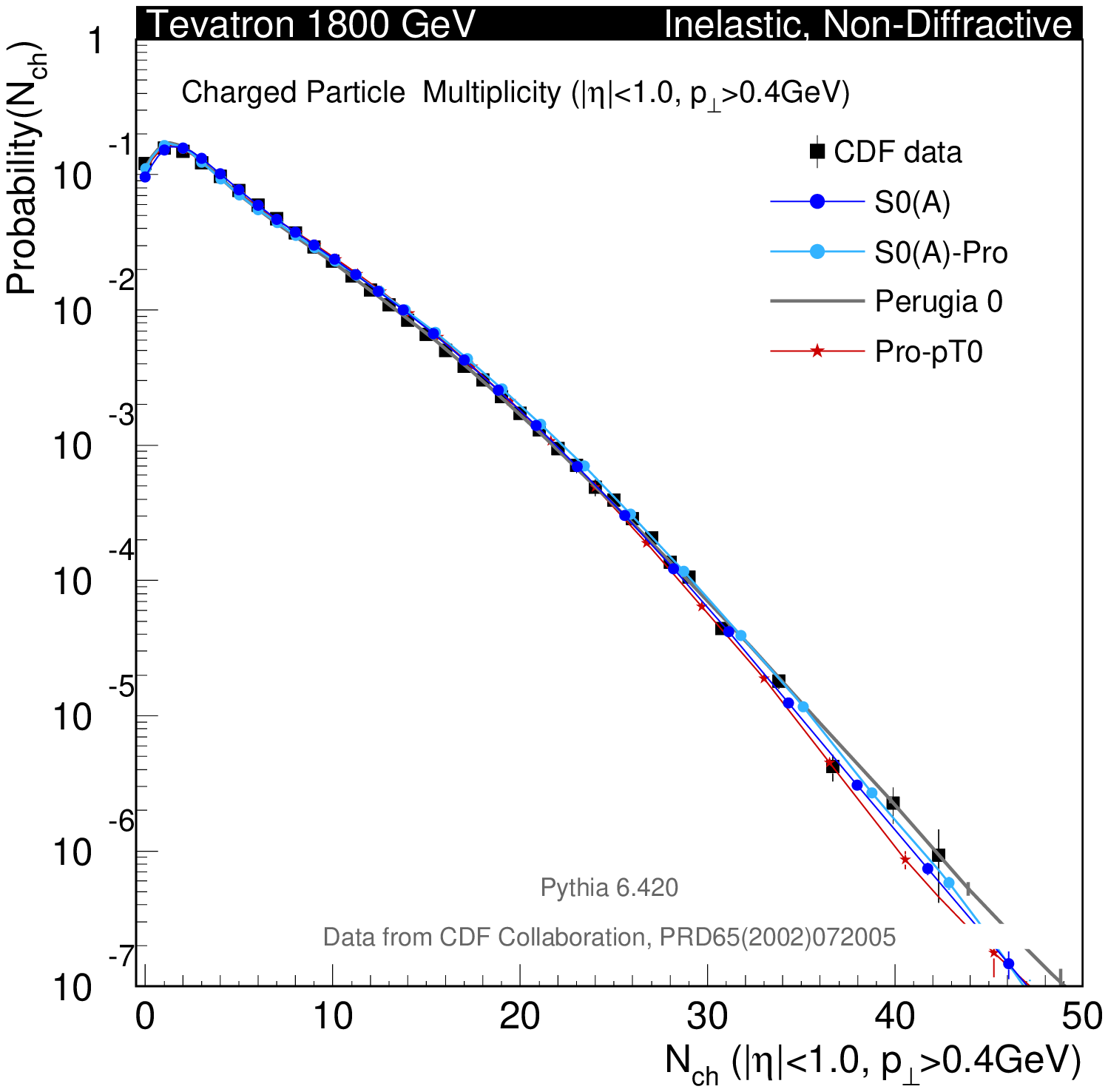}\hspace*{-5mm}
\includegraphics*[scale=0.34]{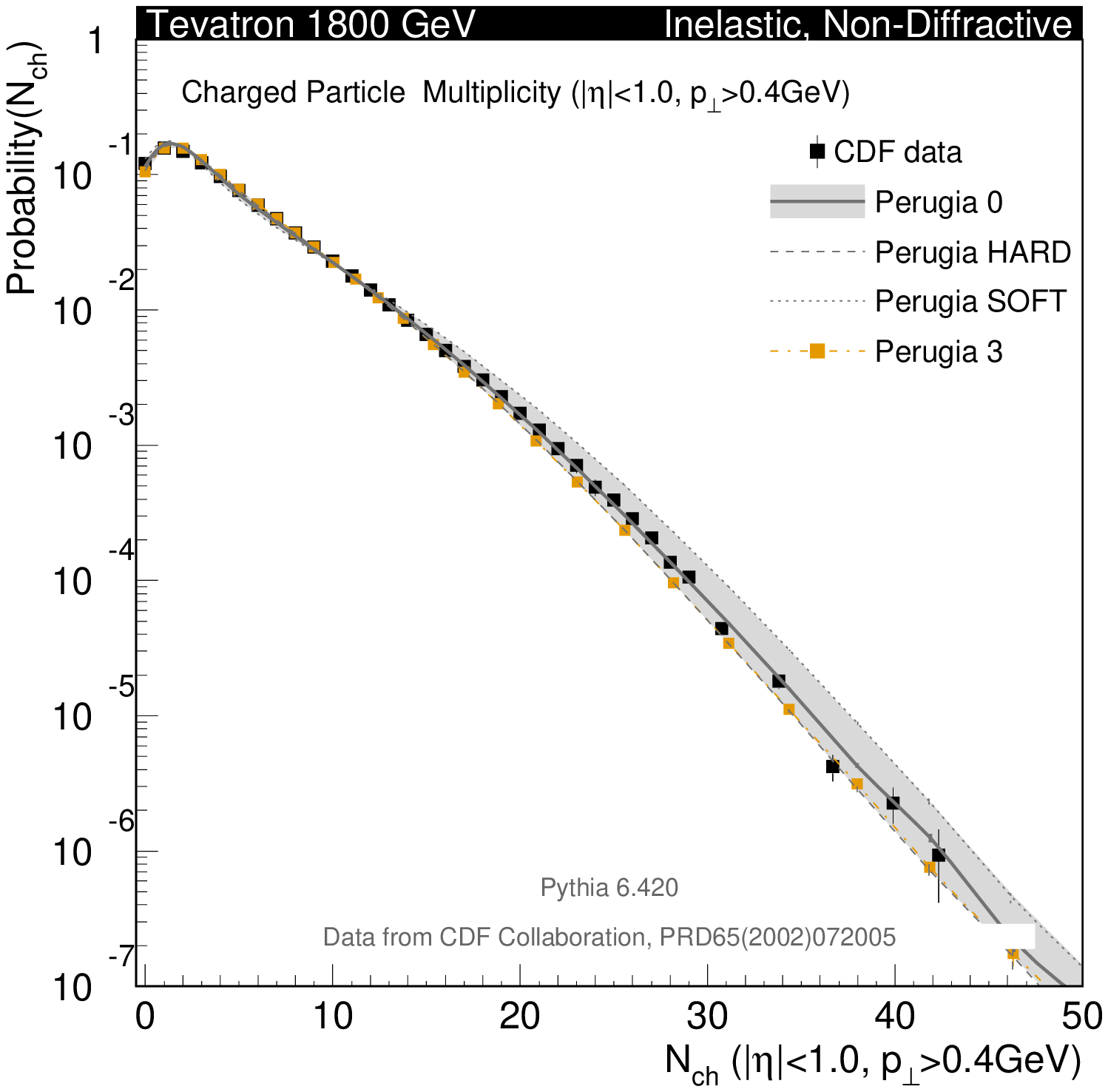}\hspace*{-8mm}\\[-3mm]
\caption{Comparisons to the CDF measurements of
 the charged track multiplicity
in minimum-bias $p\bar{p}$ collisions 
at 630 GeV (top row) and at 1800 GeV (bottom row).  
{\sl Left:} a representative
 selection of models. {\sl Center:} different tunes of the new
 framework. {\sl Right:} the
 range spanned by the  main Perugia variations. 
Results with more tunes can be found at 
 \cite{lhplots}. 
\label{fig:tevatronNCH}}
\end{center}
\end{figure}
Secondly, as mentioned above, we here include data from different
colliders at different energies, in an attempt to fix the energy
scaling better. Like Rick Field, we find that the default
energy scaling behaviour in \textsc{Pythia} results in the overall activity
growing too fast with collider energy. This can be mitigated by increasing the
dependence of the MPI infrared cutoff on collider energy. For Tune A,
Rick Field increased the power of this dependence from $\propto
E_\mrm{cm}^{0.16}$ (the default, see \cite{Sjostrand:2006za}) to $\propto
E_\mrm{cm}^{0.25}$. The
Perugia tunes incorporate a large range of values, between $0.22$ and
$0.32$, with Perugia 0 using $0.26$, i.e., very close to
the Tune A value. Note that the default was originally motivated
by the scaling of the total cross section, which grows like  $\propto
{(E_\mrm{cm}^{2})}^{0.08}$. It therefore seems that at least in the
current models, the colour screening / infrared cutoff of the
individual multi-parton interactions needs to scale
significantly faster than the total cross section. A discussion of 
whether this tendency could be given a meaningful physical
interpretation (e.g., in terms of low-$x$, saturation, or
unitarisation effects) is beyond the scope of this contribution.

As evident from fig.~\ref{fig:tevatronNCH}, the Perugia tunes all describe
the Tevatron 
$N_\mrm{ch}$ distributions at 630 (top) and 1800 (bottom) GeV within
an acceptable margin. Note that the charged track definition is here 
$\pT{}>0.4$ GeV, $|\eta|<1.0$, and particles with $c\tau\ge10$mm
treated as stable. To highlight the difference in the scaling, 
the middle plot shows both Tune S0 and Tune S0A
at 630 GeV. These are identical at 1800 GeV and 
only differ by the energy scaling, with S0 using the default scaling 
mentioned above and S0A using the Tune A value. It is mainly the
comparative failure of S0 with the default scaling to describe the 630
GeV data on the top middle plot in  fig.~\ref{fig:tevatronNCH}
that drives the choice of a slower-than-default pace of the
energy scaling of the activity (equivalent to a higher scaling power
of the infrared cutoff, as discussed above).  

\begin{figure}[t]
\begin{center}
\hspace*{-2mm}
\includegraphics*[scale=0.34]{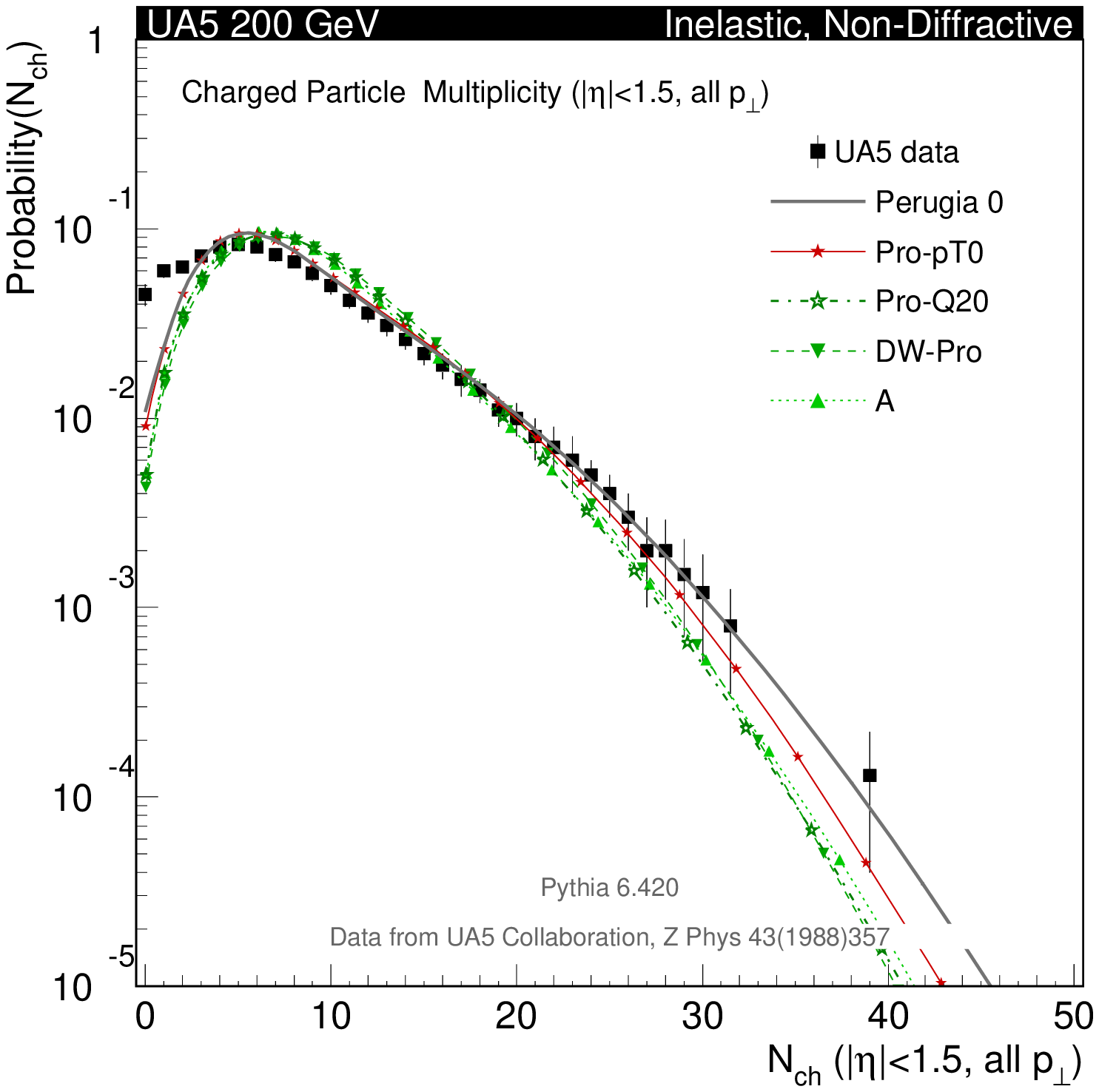}\hspace*{-5mm}
\includegraphics*[scale=0.34]{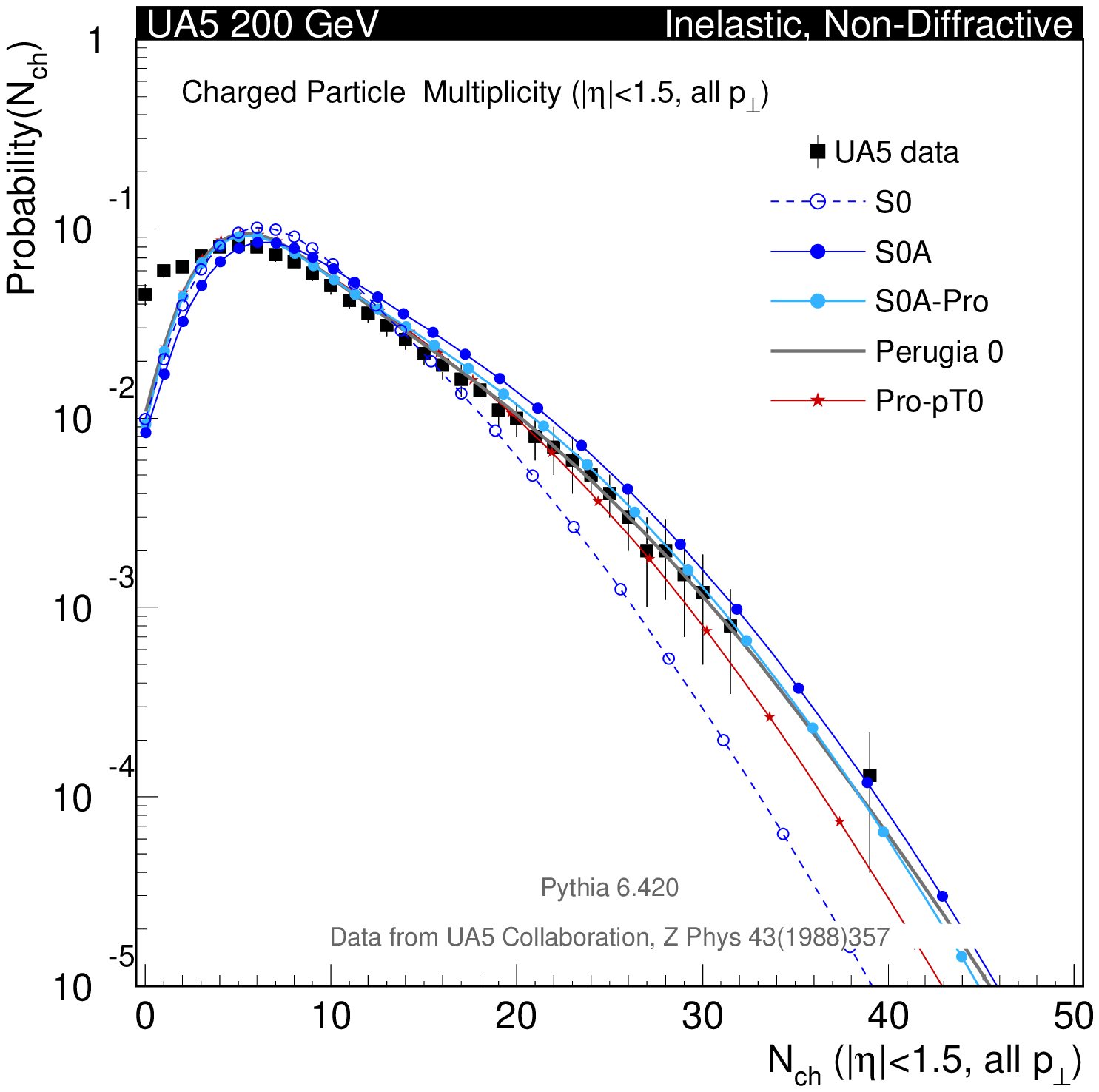}\hspace*{-5mm}
\includegraphics*[scale=0.34]{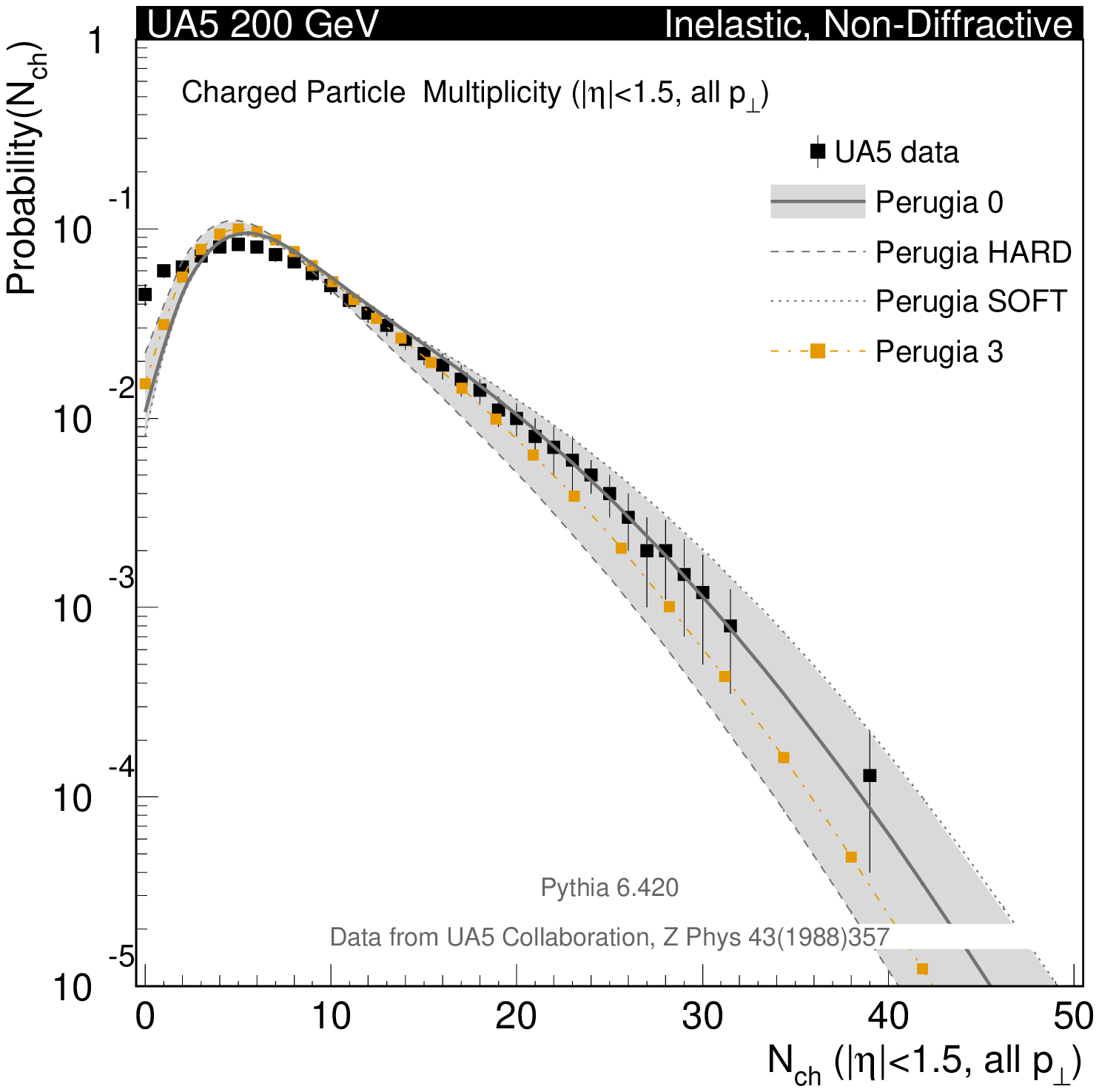}\hspace*{-8mm}\\[-5mm]
\hspace*{-2mm}
\includegraphics*[scale=0.34]{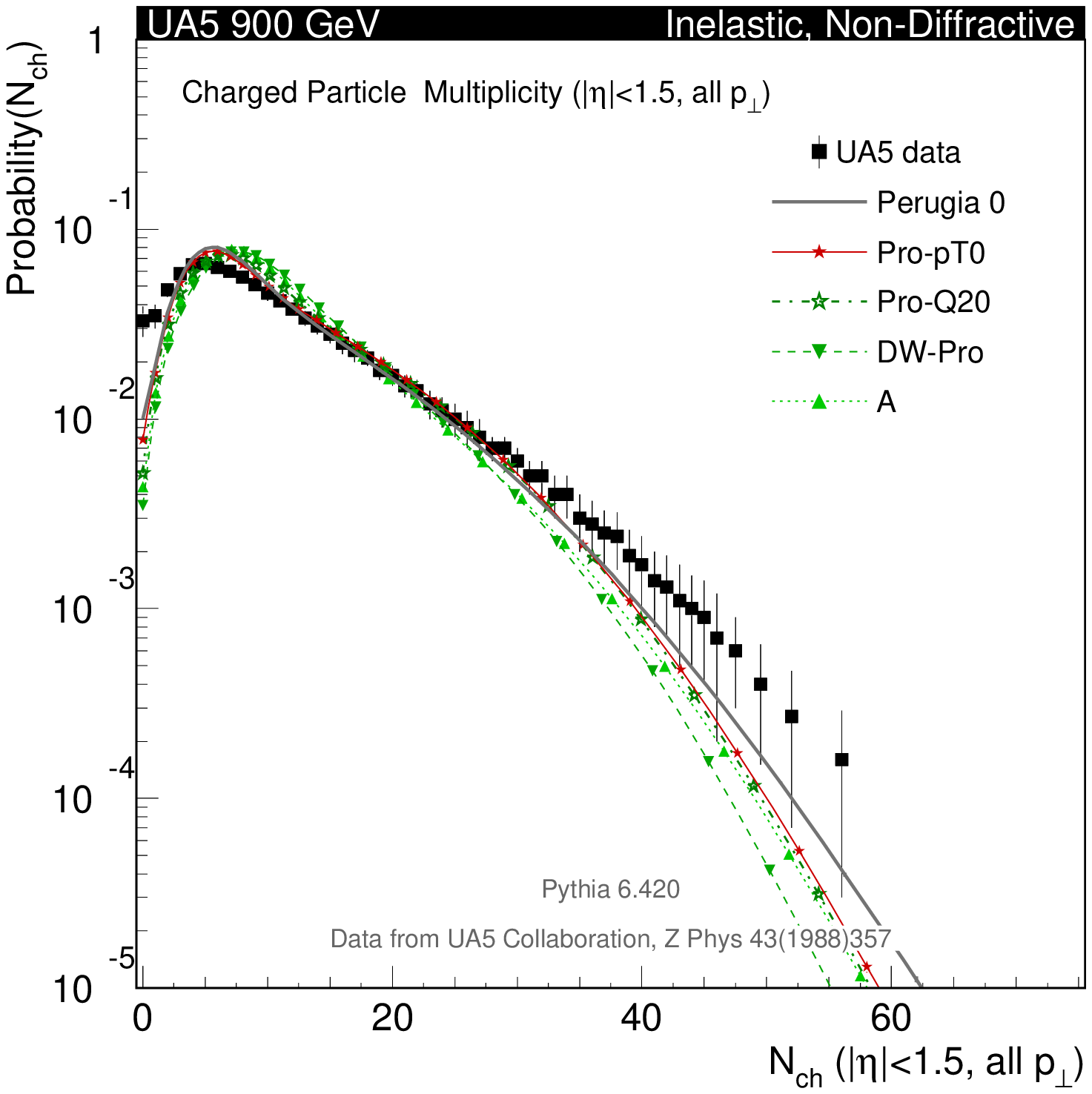}\hspace*{-5mm}
\includegraphics*[scale=0.34]{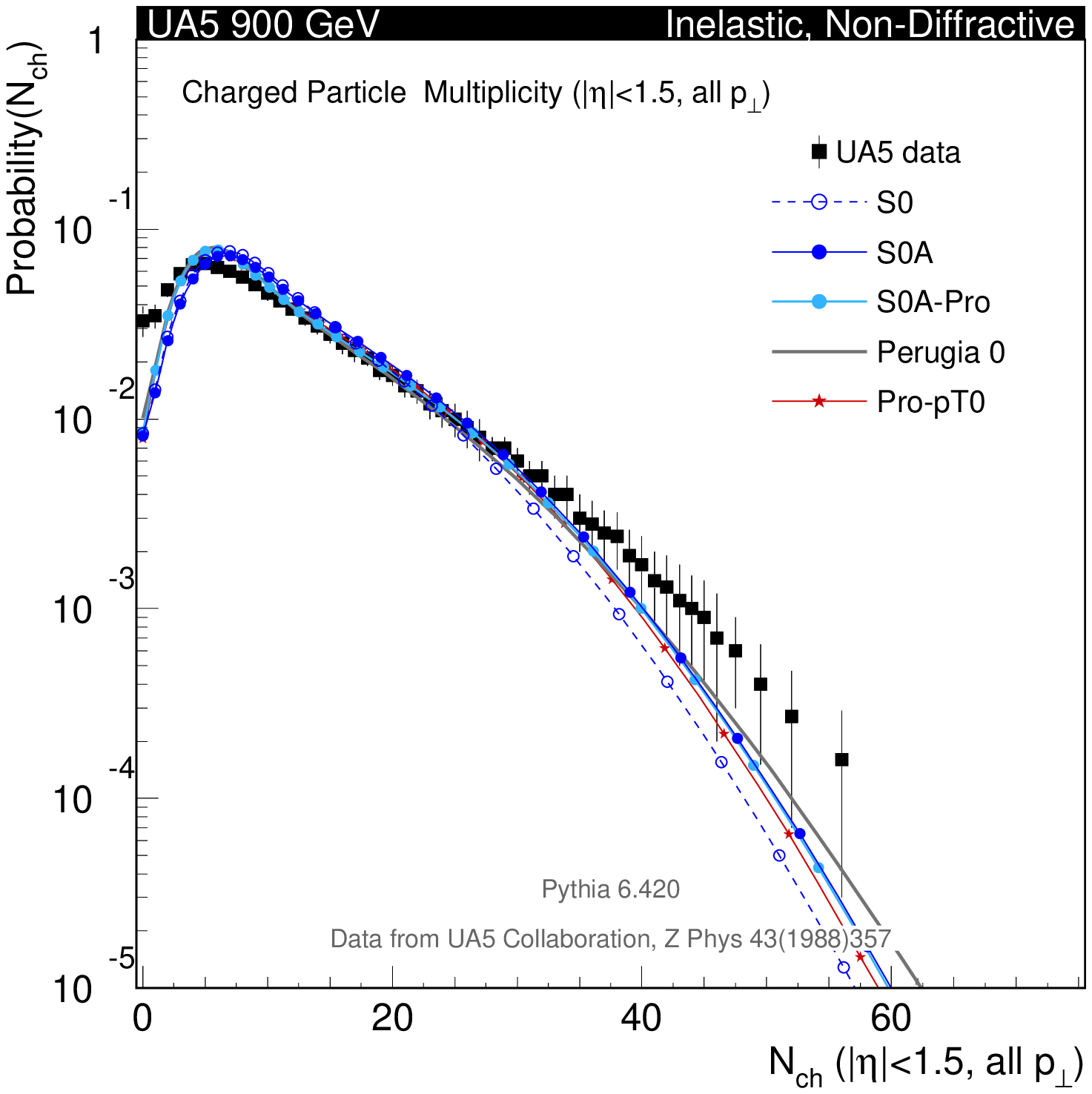}\hspace*{-5mm}
\includegraphics*[scale=0.34]{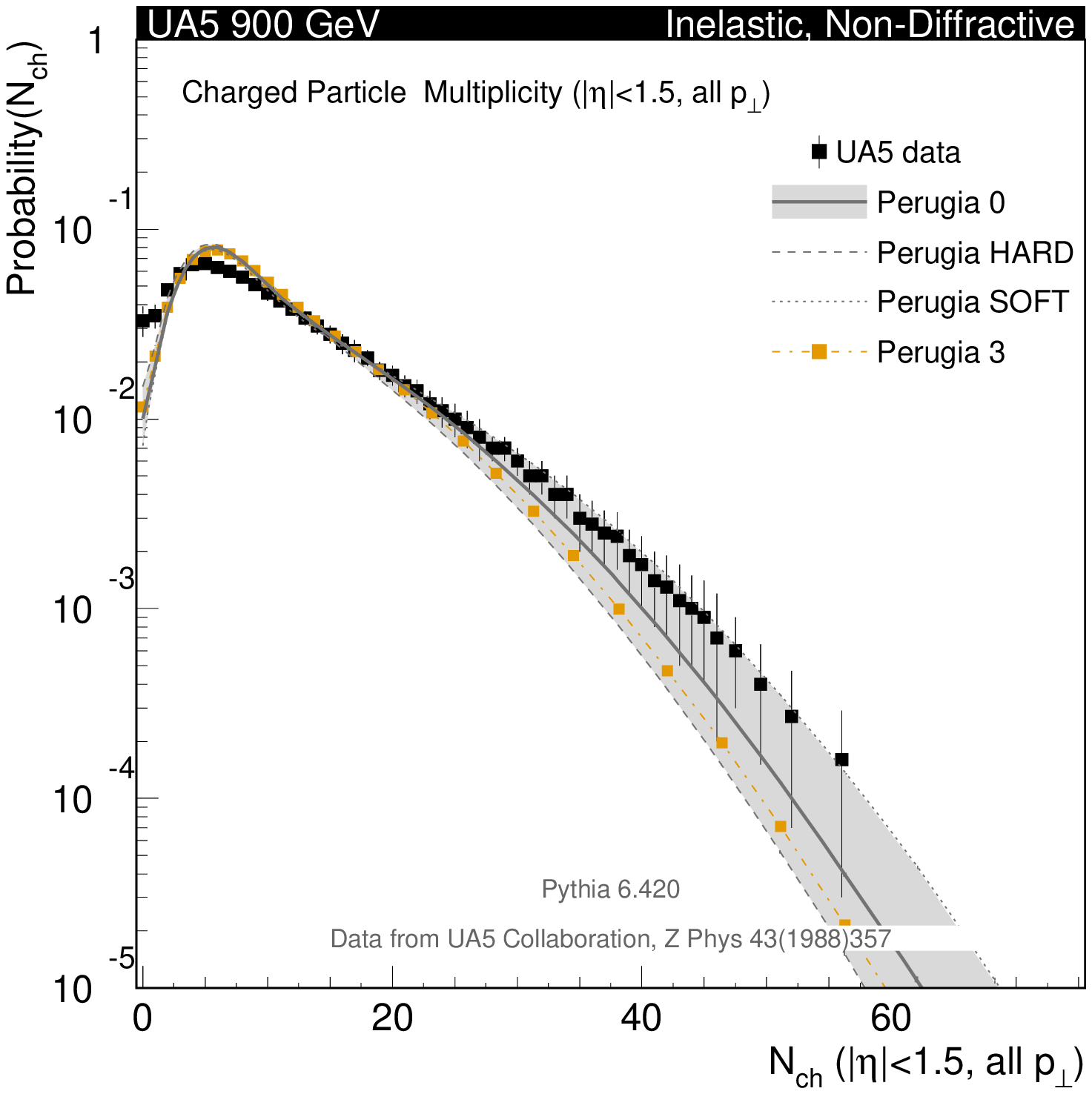}\hspace*{-8mm}\\[-3mm]
\caption{Comparisons to the UA5 measurements of
 the charged track multiplicity
in minimum-bias $p\bar{p}$ collisions 
at 200 GeV (top row) and at 900 GeV (bottom row).  
{\sl Left:} a representative
 selection of models. {\sl Center:} different tunes of the new
 framework. {\sl Right:} the
 range spanned by the  main Perugia variations.
More results can be found at  \cite{lhplots}. 
\label{fig:ua5NCH}}
\end{center}
\end{figure}
A similar comparison to UA5 data at two different energies, but now
in a slightly larger $\eta$ region and including all $\pT{}$ is shown in
fig.~\ref{fig:ua5NCH}. Since the data here includes all $\pT{}$, the
theoretical models have been allowed to deviate slightly more from the
data than for the Tevatron and the first few bins were ignored, 
to partly reflect uncertainties associated with the production of very soft
particles. 

The good news, from the point of view of LHC physics, is that even the
most extreme Perugia variants need to have a more slowly growing
activity than the default. Thus, their extrapolations to the LHC
produce \emph{less} underlying event than those of their predecessors
that used the default scaling, such as S0, DWT, or ATLAS-DC2/Rome.

\begin{figure}[t]
\begin{center}\hspace*{-2mm}
\includegraphics*[scale=0.34]{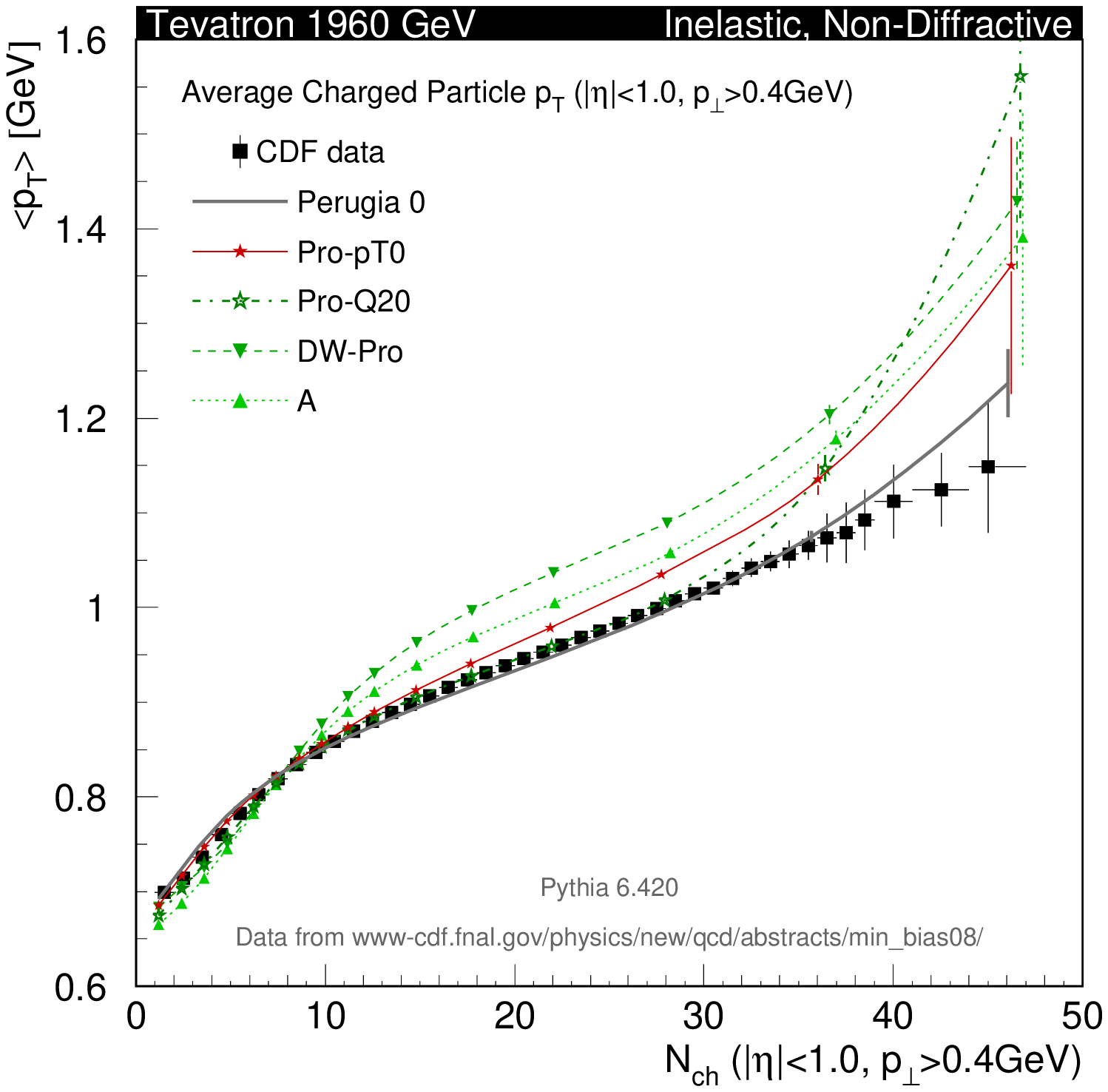}\hspace*{-5mm}
\includegraphics*[scale=0.34]{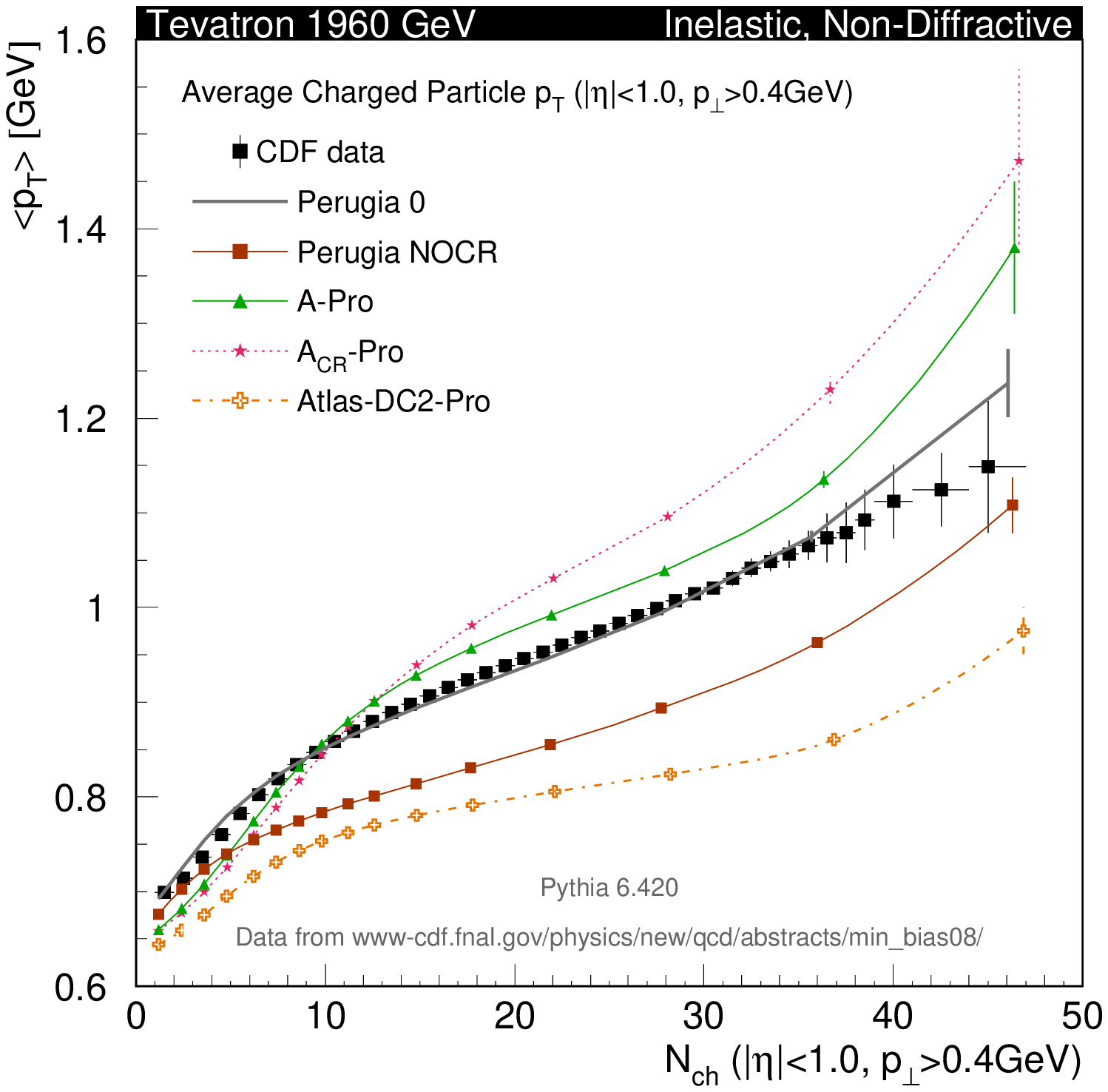}\hspace*{-5mm}
\includegraphics*[scale=0.34]{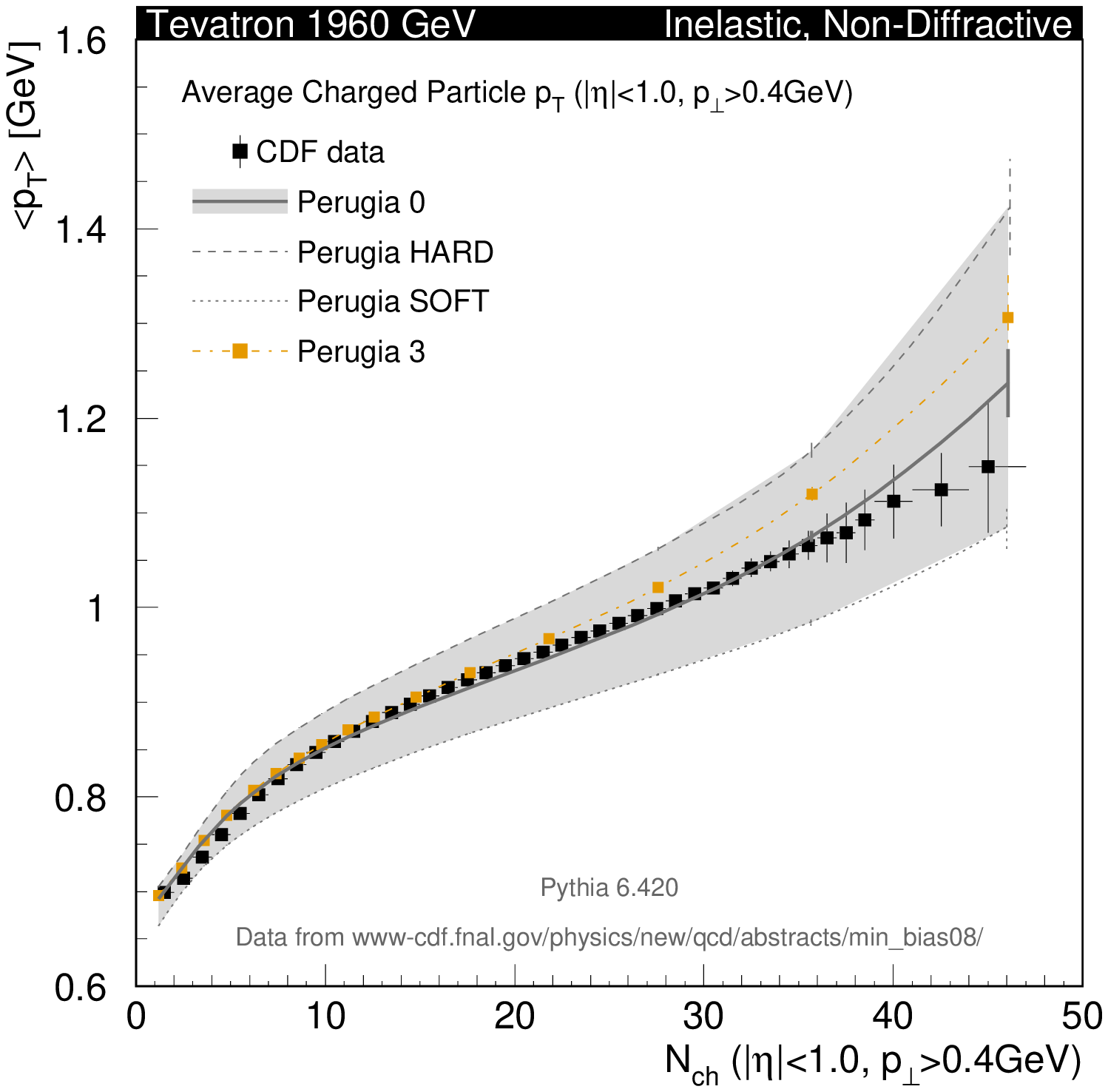}\hspace*{-8mm}\\[-3mm]
\caption{Comparisons to the CDF Run II measurement of
 the average track $\pT{}$ as a function of track multiplicity 
in min-bias $p\bar{p}$ collisions.  {\sl Left:} a representative
 selection of models. {\sl Center:} the impact of varying models
 of color (re-)connections on this distribution. {\sl Right:} the
 range spanned by the  main Perugia variations. The SOFT and HARD
 variations were here allowed to deviate by significantly 
more than the statistical precision due to the high sensitivity of the
 distribution and the large theoretical uncertainties. 
Results with more tunes can be found at 
 \cite{lhplots}. 
\label{fig:tevatronAVG}}
\end{center}
\end{figure}
Thirdly, while the charged particle $\pT{}$ spectrum (see
\cite[dN/dpT]{lhplots}) and $N_{\mrm{ch}}$ distribution in Tune A was
in almost perfect 
agreement with Tevatron min-bias data, the high-multiplicity behaviour
of the $\left<N_\mrm{ch}\right>(\pT{})$ distribution was slightly too
high \cite{moggi}. This slight discrepancy carried over to the S0 family of tunes
of the new framework, since these were tuned to Tune A, in the absence of
published data. Fortunately, CDF data has now been made publicly
available \cite{moggi}, and hence it was possible to take the actual
data into consideration for the Perugia tunes, resulting in somewhat softer
particle spectra in high-multiplicity events,
cf.~fig.~\ref{fig:tevatronAVG}. Note that this distribution is highly
sensitive to the colour structure of the events, as emphasized in
\cite{Sjostrand:1987su,Sandhoff:2005jh,Skands:2007zg,Wicke:2008iz}. 

Finally, the old
framework did not include showering off the MPI
in- and out-states\footnote{It did, of course, include showers off the
  primary interaction. S.~Mrenna has since implemented FSR off the MPI 
as an additional option in that framework, 
but tunes using that option have not yet been
made.}. The new framework does include such showers, which furnishes
an additional fluctuating physics component. Relatively speaking, 
the new framework therefore  needs \emph{less}
fluctuations from other sources in order to describe the
same data. This is reflected in the tunes of the new framework
generally having a less lumpy proton (smoother proton transverse
density distributions) 
and fewer total numbers of MPI than the old one. We included
illustrations of this in a special ``theory'' section of the web
plots, cf.~\cite[Theory Plots]{lhplots} and \cite[Fig.~4]{Skands:2007zz}. 

The showers off the MPI also lead to a greater
degree of decorrelation and $\pT{}$ imbalance 
between the minijets produced by the
underlying event, in contrast to the old framework where these 
remained almost exactly balanced and back-to-back. This should show up
in minijet $\Delta\phi_{jj}$ and/or $\Delta R_{jj}$ distributions sensitive to the
underlying event, such as in $Z/W$+jets with low $\pT{}$ cuts on the
additional jets. 

Further, 
since showers tend to produce shorter-range correlations than MPI, the
new tunes also exhibit smaller long-range correlations than the old
models. I.e., if 
there is a large fluctuation in one end of the detector, it is
\emph{less} likely in the new models that there is a large fluctuation in
the same direction in the other end of the detector. The impact of
this, if any, on the overall modeling and correction procedures derived from
it, has not yet been studied. At the very least it 
 furnishes a systematic difference between the models. For brevity, we
do not include the plots here but refer to the web \cite[FB
  Correlation]{lhplots} and to the original \textsc{Pythia} MPI paper 
for a definition and comparable plots \cite{Sjostrand:1987su}.  

\section{Tune-by-Tune Descriptions}

The starting point for all the Perugia tunes, apart from Perugia NOCR,
was S0(A)-Pro, i.e., the original tunes S0 and S0A, revamped to
include the Professor tuning of flavour and fragmentation parameters
to LEP data \cite{HoethProc}. The starting point for Perugia NOCR was
NOCR-Pro. From these starting points, the main hadron collider parameters
were retuned to better describe the above mentioned data sets. An
overview of the tuned parameters and their values is given in table
\ref{tab:parameters}.

\begin{table}[tp]
\begin{center}
\begin{tabular}{lc|r|rrrrrrr|}
Parameter & Type & S0A-Pro & P-0 & P-HARD & P-SOFT & P-3 & P-NOCR & P-X & P-6 \\
\toprule
\ttt{MSTP(51)} & PDF &   7 &   7 &      7 &      7 &    7 &      7 &
20650 & 10042 \\
\ttt{MSTP(52)} & PDF &   1 &   1 &      1 &      1 &    1 &      1 &   
2     &     2 \\
\cmidrule{1-10}
\ttt{MSTP(64)} & ISR &   2 &   3 &      3 &      2 &    3 &      3 &    
3     &     3 \\
\ttt{PARP(64)} & ISR & 1.0 & 1.0 &   0.25 &    2.0 &  1.0 &    1.0 & 
2.0   &    1.0\\
\ttt{MSTP(67)} & ISR &   2 &   2 &      2 &      2 &    2 &      2 &
2     &      2\\
\ttt{PARP(67)} & ISR & 4.0 & 1.0 &    4.0 &    0.5 &  1.0 &    1.0 & 
1.0   &    1.0\\
\ttt{MSTP(70)} & ISR &   2 &   2 &      0 &      1 &    0 &      2 &
2     &      2\\
\ttt{PARP(62)} & ISR &   - &   - &   1.25 &      - & 1.25 &      - &
-     &      -\\
\ttt{PARP(81)} & ISR &   - &   - &      - &    1.5 &    - &      - &
-     &      -\\
\ttt{MSTP(72)} & ISR &   0 &   1 &      1 &      0 &    2 &      1 &
1     &      1\\
\cmidrule{1-10}
\ttt{PARP(71)} & FSR & 4.0 & 2.0 &    4.0 &    1.0 &  2.0 &    2.0 &
2.0   &    2.0\\
\ttt{PARJ(81)} & FSR &0.257&0.257&   0.3  &   0.2 & 0.257 & 0.257 &
0.257 &  0.257  \\
\ttt{PARJ(82)} & FSR & 0.8 & 0.8 &   0.8 &    0.8 &  0.8 &    0.8 & 
0.8 &   0.8 \\
\cmidrule{1-10}
\ttt{MSTP(81)} &  UE &  21 &  21 &     21 &     21 &   21 &     21 &
21    &     21\\
\ttt{PARP(82)} &  UE &1.85 & 2.0 &    2.3 &    1.9 &  2.2 &   1.95 &   
2.2   &   1.95\\
\ttt{PARP(89)} &  UE &1800 &1800 &   1800 &   1800 & 1800 &   1800 &
1800  &   1800\\  
\ttt{PARP(90)} &  UE &0.25 &0.26 &   0.30 &   0.24 & 0.32 &   0.24 &
0.23  &  0.22\\
\ttt{MSTP(82)} &  UE &   5 &   5 &      5 &      5 &    5 &      5 &
5     &     5\\
\ttt{PARP(83)} &  UE & 1.6 & 1.7 &    1.7 &    1.5 &  1.7 &    1.8 &
1.7   &   1.7\\
\cmidrule{1-10}
\ttt{MSTP(88)} &  BR &   0 &   0 &      0 &      0 &    0 &      0 &
0     &     0\\
\ttt{PARP(79)} &  BR & 2.0 & 2.0 &    2.0 &    2.0 &  2.0 &    2.0 &
2.0   &   2.0\\
\ttt{PARP(80)} &  BR &0.01 &0.05 &   0.01 &   0.05 & 0.03 &   0.01 &
0.05  &  0.05\\
\ttt{MSTP(91)} &  BR &   1 &   1 &      1 &      1 &    1 &      1 &
1     &     1\\
\ttt{PARP(91)} &  BR & 2.0 & 2.0 &    1.0 &    2.0 &  1.5 &    2.0 &
2.0   &   2.0\\
\ttt{PARP(93)} &  BR &10.0 & 10.0&   10.0 &   10.0 & 10.0 &  10.0 &
10.0  &  10.0\\
\cmidrule{1-10}
\ttt{MSTP(95)} &  CR &   6 &   6 &      6 &      6 &    6 &     6 &
6     &     6\\
\ttt{PARP(78)} &  CR & 0.2 &0.33 &   0.37 &    0.15& 0.35 &   0.0 &
0.33  &  0.33\\ 
\ttt{PARP(77)} &  CR & 0.0 & 0.9 &   0.4  &    0.5 & 0.6  &   0.0 &
0.9   &   0.9\\
\cmidrule{1-10}
\ttt{MSTJ(11)} & HAD &   5 &   5 &     5  &      5 &    5 &     5 &
5     &     5\\
\ttt{PARJ(21)} & HAD &0.313&0.313 &  0.34 &   0.28 & 0.313& 0.313 &
0.313 &  0.313\\
\ttt{PARJ(41)} & HAD & 0.49& 0.49 &  0.49 &   0.49 & 0.49 &  0.49 &
0.49  &   0.49\\
\ttt{PARJ(42)} & HAD & 1.2& 1.2 &  1.2   &   1.2 & 1.2 &  1.2 &
1.2 &   1.2\\
\ttt{PARJ(46)} & HAD & 1.0& 1.0 &  1.0   &   1.0 & 1.0 &  1.0 &
1.0 &   1.0\\  
\ttt{PARJ(47)} & HAD & 1.0& 1.0 &  1.0   &   1.0 & 1.0 &  1.0 &
1.0 &   1.0\\
\bottomrule
\end{tabular}
\caption{Parameters of the Perugia tunes, omitting the LEP flavour
  parameters tuned by Professor \cite{HoethProc} (common to all 
   the ``Pro'' and ``Perugia'' tunes). The starting point, S0A-Pro, is
   shown for reference. (BR stands for Beam Remnants
   and CR stands for Colour Reconnections.)\label{tab:parameters}}
\end{center}
\end{table}

\paragraph{Perugia 0 (320): } 
Uses $\Lambda_\mrm{CMW}$ instead of $\Lambda_{\overline{\mrm{MS}}}$,
which results in near-perfect agreement with the Drell-Yan $p_\perp$
spectrum, both in the tail and in the peak,
cf.~fig.~\ref{fig:tevatronDY}, middle plot.  
Also has slightly less colour reconnections, especially among 
high-$\pT{}$ string pieces, which improves the agreement both with 
the $\left<\pT{}\right>(N_\mrm{ch})$ distribution and with the
high-$\pT{}$ tail 
of charged particle $\pT{}$ spectra, cf~\cite[dN/dpT (tail)]{lhplots}). 
Compared to S0A-Pro, this tune also has slightly more beam-remnant
breakup  (more baryon number transport), mostly in order to explore this
possibility than due to any necessity of tuning.
Without further changes, these modifications would lead to a greatly
increased average multiplicity as well as larger multiplicity
fluctuations. To keep the total multiplicity unchanged,
cf.~the solid grey curves labeled ``Perugia 0'' on the plots 
in the top row of fig.~\ref{fig:tevatronNCH}, the
changes above were accompanied by an increase in the MPI infrared
cutoff, which decreases the overall MPI-associated activity, and 
by a slightly smoother proton mass profile, which decreases the
fluctuations. Finally, the energy scaling is closer to that of S0A
than to the default one used for S0, cf.~the middle panes in 
figs.~\ref{fig:tevatronNCH} and \ref{fig:ua5NCH}.  
\paragraph{Perugia HARD (321): } Variant of Perugia 0 which has a
higher amount of activity from perturbative physics and
counter-balances that partly by having less particle production from
nonperturbative sources. Thus, the $\Lambda_{{\mrm{CMW}}}$ 
value is used for ISR, together with a
renormalisation scale for ISR of $\mu_R=\frac12\pT{}$, yielding a
comparatively hard Drell-Yan $\pT{}$ spectrum,
cf.~the dashed curve labeled ``HARD'' in the 
right pane of fig.~\ref{fig:tevatronDY}. It also has a
slightly larger phase space 
for both ISR and FSR, uses higher-than-nominal values for
FSR, and has a slightly harder hadronisation. To partly counter-balance
these choices, it has less ``primordial $k_T$'', a higher infrared
cutoff for the MPI, and more active color reconnections,
yielding a comparatively high curve for
$\left<\pT{}\right>(N_\mrm{ch})$, cf.~fig.~\ref{fig:tevatronAVG}. 
\paragraph{Perugia SOFT (322): } Variant of Perugia 0 which has a
lower amount of activity from perturbative physics and makes up for it
partly by adding more particle production from nonperturbative sources. 
Thus, the $\Lambda_{\overline{\mrm{MS}}}$ value is used for ISR, 
together with a
renormalisation scale of $\mu_R=\sqrt{2}\pT{}$, yielding a
comparatively soft Drell-Yan $\pT{}$ spectrum,
cf.~the dotted curve labeled ``SOFT'' in the right pane of 
fig.~\ref{fig:tevatronDY}. It also has a slightly smaller phase space
for both ISR and FSR, uses lower-than-nominal values for
FSR, and has a slightly softer hadronisation. To partly counter-balance
these choices, it has a more 
sharply peaked proton mass distribution, a more active beam remnant
fragmentation (lots of baryon transport), a slightly lower infrared
cutoff for the MPI, and slightly less active color reconnections,
yielding a comparatively low curve for
$\left<\pT{}\right>(N_\mrm{ch})$, cf.~fig.~\ref{fig:tevatronAVG}. 
\paragraph{Perugia 3 (323): } Variant of Perugia 0 which has a
different balance between MPI and ISR and a different energy
scaling. Instead of a smooth dampening of ISR all the way to zero
$\pT{}$, this tune uses a sharp cutoff at 1.25 GeV, which produces
a slightly harder ISR spectrum. The additional ISR activity is
counter-balanced by a higher infrared MPI cutoff. Since the ISR cutoff
is independent of the collider CM energy in this tune, 
the multiplicity would nominally evolve very rapidly with energy. To 
offset this, the MPI cutoff itself must scale very quickly, hence
this tune has a very large value of the scaling power of that
cutoff. This leads to an interesting systematic difference in the
scaling behaviour, with ISR becoming an increasingly more important 
source of particle production as the energy increases in this tune,
relative to Perugia 0. 
\paragraph{Perugia NOCR (324): } An update of NOCR-Pro that attempts
to fit the data sets as well as possible, without invoking any
explicit colour reconnections. Can reach an acceptable agreement with
most distributions, except for the $\left<\pT{}\right>(N_\mrm{ch})$
one, cf.~fig.~\ref{fig:tevatronAVG}. 
\paragraph{Perugia X (325): } A Variant of Perugia 0 which uses the
MRST LO* PDF set \cite{Sherstnev:2007nd}. 
Due to the increased gluon densities, a slightly 
lower ISR renormalisation scale and a higher MPI cutoff than for
Perugia 0 is used. Note that, since we are not yet sure the
implications of using LO* for the MPI interactions have been fully
understood, this tune should be considered experimental for the time
being. See \cite[Perugia PDFs]{lhplots} for distributions.
\paragraph{Perugia 6 (326): } A Variant of Perugia 0 which uses the
CTEQ6L1 PDF set \cite{Pumplin:2002vw}. 
Identical to Perugia 0 in all other respects, except
for a slightly lower MPI infrared cutoff at the Tevatron and a
lower scaling power of the MPI infrared cutoff. See \cite[Perugia PDFs]{lhplots} for
distributions.

\section{Extrapolation to the LHC}
Part of the motivation for updating the S0 family of tunes was
specifically to improve the constraints on the 
energy scaling to come up with tunes
that extrapolate more reliably to the LHC. This is not to
say that the uncertainty is still not large, but as mentioned above,
it does seem that, e.g., the default \textsc{Pythia} scaling has by
now been convincingly ruled out, and so this is naturally reflected in
the updated parameters. 

\begin{figure}[t]
\begin{center}\hspace*{-2mm}
\includegraphics*[scale=0.34]{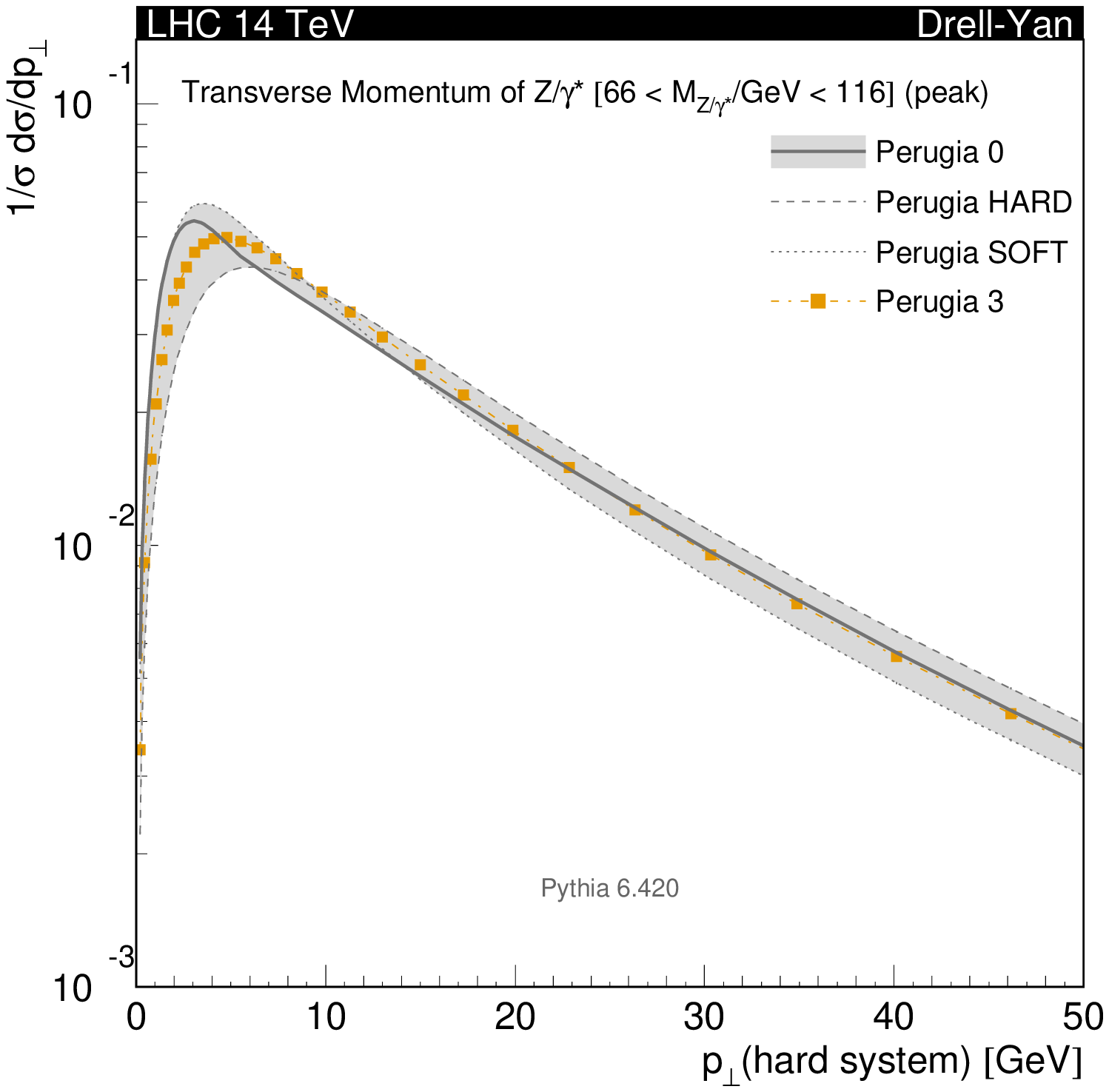}\hspace*{-5mm}
\includegraphics*[scale=0.34]{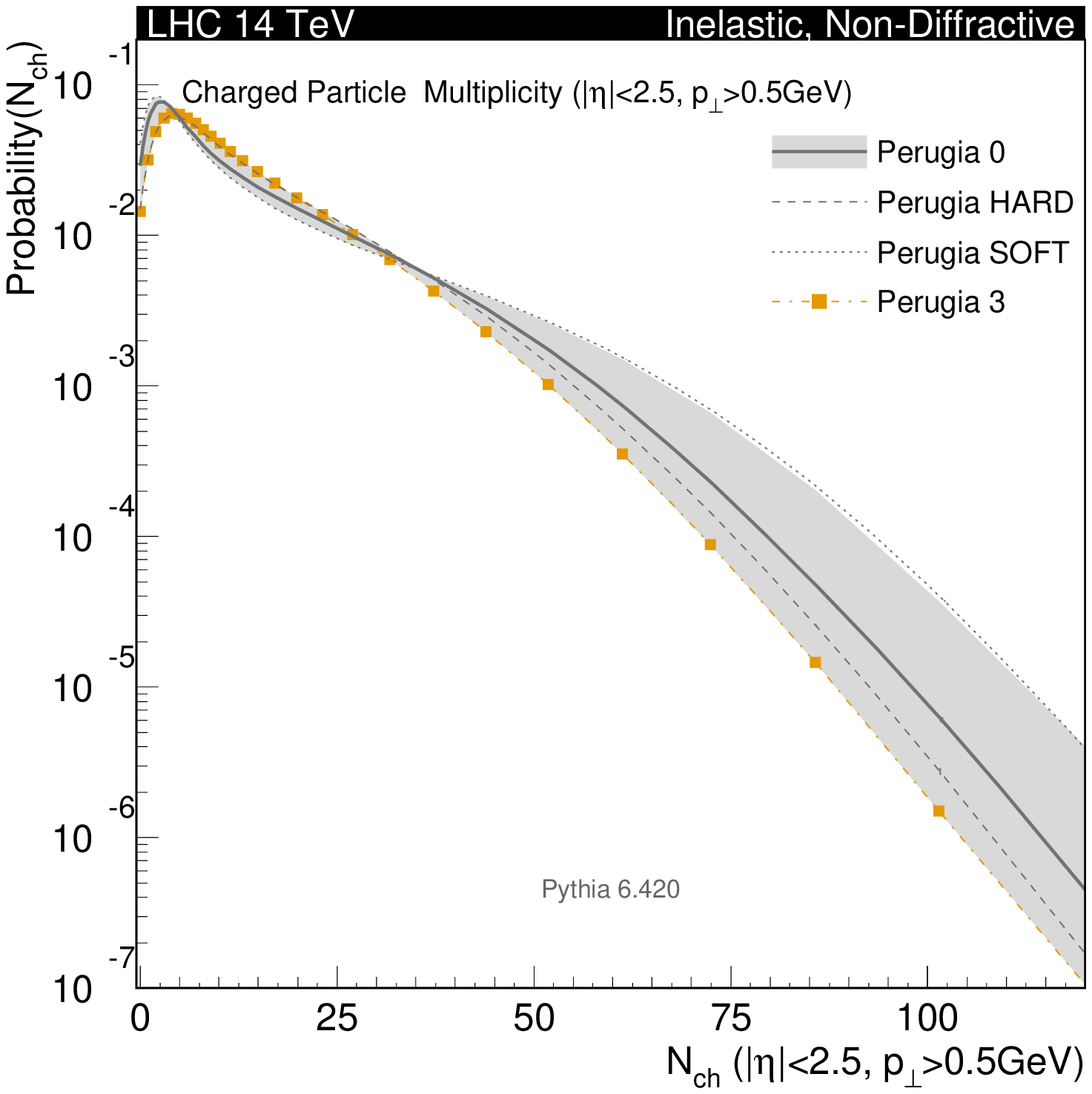}\hspace*{-5mm}
\includegraphics*[scale=0.34]{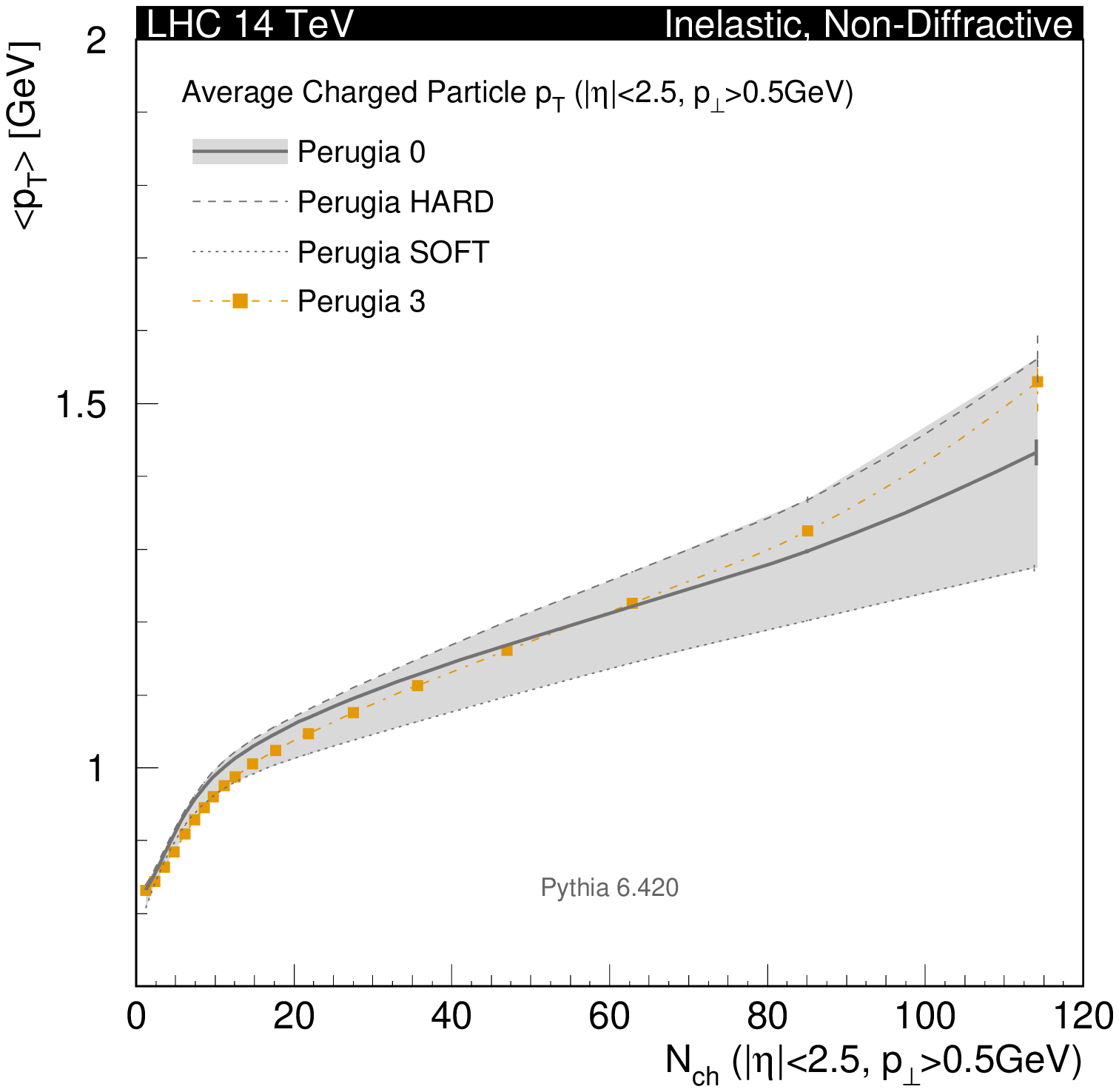}\hspace*{-8mm}\\[-4mm]
\caption{Perugia ``predictions'' for the $\pT{}$ of Drell-Yan pairs (left),
  the charged track multiplicity in min-bias (center), and the average
  track $\pT{}$ in min-bias (right) at the LHC. See
  \cite{lhplots} for additional plots.
\label{fig:LHC}}
\end{center}
\end{figure}
Fig.~\ref{fig:LHC} contains predictions for the Drell-Yan
$\pT{}$ distribution (using the CDF cuts), the charged track
multiplicity distribution in minimum-bias collisions, and the average
track $\pT{}$ as a function of multiplicity at 14 TeV, for the
central, hard, soft, and ``3'' variations of the Perugia tunes. We
hope this helps to give a feeling for the kind of ranges spanned by
the Perugia tunes (the PDF variations give almost identical results to
Perugia 0 for these distributions). A full set of plots illustrating
the extrapolations to the 
LHC for both the central region $|\eta|<2.5$ as well as the region 
$1.8<\eta<4.9$ covered by LHCb can be found on the web
\cite{lhplots}. 

However, in addition to these plots, 
we thought it would be interesting to make at
least one set of numerical predictions for an infrared sensitive
quantity that could be tested with the very earliest LHC data. 
We therefore used
the Perugia tunes and their variations to get an estimate for the mean 
multiplicity of charged tracks in (inelastic, nondiffractive) 
minimum-bias $pp$ collisions at 10 and 14 TeV. The Perugia variations
indicate an uncertainty of order 15\% or less on the central
values, which is probably an underestimate, due to the limited nature
of the models. Nonetheless, having spent a significant amount of
effort in making these estimates, given in tab.~\ref{tab:predictions},
we intend to stick by them until proved wrong. The acknowledgments
therefore contain a recognition of a bet to that effect. 

\begin{table}[t]
\begin{center}
\begin{tabular}{lrrrrrr}
\multicolumn{6}{c}{\sl Predictions for Mean Densities of Charged Tracks} \\[1mm]\toprule\\[-3mm]
           & $\displaystyle\frac{\left<N_\mrm{ch}\right>|_{N_{\mrm{ch}}\ge 0}}{\Delta\eta\Delta\phi}$ 
           & $\displaystyle\frac{\left<N_\mrm{ch}\right>|_{N_{\mrm{ch}}\ge 1}}{\Delta\eta\Delta\phi}$
           & $\displaystyle\frac{\left<N_\mrm{ch}\right>|_{N_{\mrm{ch}}\ge 2}}{\Delta\eta\Delta\phi}$
           & $\displaystyle\frac{\left<N_\mrm{ch}\right>|_{N_{\mrm{ch}}\ge 3}}{\Delta\eta\Delta\phi}$
           & $\displaystyle\frac{\left<N_\mrm{ch}\right>|_{N_{\mrm{ch}}\ge 4}}{\Delta\eta\Delta\phi}$
\\[4mm]LHC 10 TeV & $0.40 \pm 0.05$ & $0.41 \pm 0.05$ & $ 0.43\pm 0.05$ &
$0.46\pm 0.06$ & $0.50\pm 0.06$
\\[1mm]LHC 14 TeV & $0.44 \pm 0.05$ & $0.45 \pm 0.06$ & $0.47 \pm 0.06$ 
& $0.51 \pm 0.06$ & $0.54 \pm 0.07$
 \\[1mm]\bottomrule
\end{tabular}
\caption{Best-guess predictions for the mean density of charged
  tracks for min-bias $pp$ collisions at two LHC energies. These
  numbers should be compared to data corrected to 100\% track finding
  efficiency for tracks with $|\eta|<2.5$ and $\pT{}>0.5$ GeV
  and 0\% efficiency outside that region. The definition of a stable
  particle was set at $c\tau \ge 10$mm (e.g., the two tracks
  from a $\Lambda^0\to p^+\pi^-$ decay were not counted). 
  The $\pm$ values represent the estimated uncertainty,
  based on the Perugia tunes. Since the lowest multiplicity
  bins may receive large corrections from elastic/diffractive
  events, it is possible that it will be easier to compare the (inelastic
  nondiffractive) theory to the first data
  with one or more of the lowest multiplicity bins excluded, 
  hence we have here recomputed the means with up to the first 4 bins
  excluded. (These predictions were first shown at the 2009 Aspen Winter
  Conference.) 
\label{tab:predictions}}
\end{center}
\end{table}

\section{Conclusions}
We have presented a set of updated parameter sets (tunes) 
for the interleaved
$\pT{}$-ordered shower and underlying-event model in \textsc{Pythia}
6.4. These parameter sets include the revisions to the fragmentation and flavour
parameters obtained by the Professor group and reported on elsewhere
in these proceedings \cite{HoethProc}. The new sets further include more
Tevatron data and more data from different collider CM energies in an
attempt to simultaneously improve the overall description at the Tevatron
data while also improving the reliability of the extrapolations to the
LHC. We have also attempted to deliver a first set of 
``tunes with uncertainty bands'', by including alternative tunes with
systematically different parameter choices. The new tunes are
available from Pythia version 6.4.20, via the routine PYTUNE. 

We note that these tunes still only included Drell-Yan and minimum-bias data
directly; leading-jet, photon+jet, and underlying-event data was not
considered explicitly. This is not expected to be a major problem due
to the good universality properties that the \textsc{Pythia} modeling
has so far exhibited, but it does mean that the performance of the
tunes on such data sets should be tested, which will hopefully happen
in the near future. 

We hope these tunes will be useful to the RHIC, Tevatron, and 
LHC communities. 
\subsubsection*{Acknowledgments}
We are grateful to the organizers of this very enjoyable workshop,
which brought people from different communities 
together, and helped us take some steps towards finding a common language. 
In combination with the writeup of this article, 
I agreed to owing Lisa Randall a bottle of champagne if 
the first published measurement of any number  in 
table \ref{tab:predictions} is outside the range given. 
Conversely, she agrees to owing me a bottle if the corresponding
number happens to be right. 

\bibliography{perugia-tunes}

\end{document}